\documentclass[a4paper,11pt]{article}

\usepackage{jheppub} 
\usepackage[T1]{fontenc}
\usepackage[usenames,dvipsnames]{xcolor}
\usepackage{graphicx}
\usepackage{hyperref}
\usepackage[capitalise]{cleveref}
\usepackage{xspace}
\usepackage[ISO]{diffcoeff}
\usepackage[normalem]{ulem}

\graphicspath{{./figures/}}

\DeclareRobustCommand{\RESNOVA}{\mbox{RES-NOVA}}
\newcommand{\cevns}{CE$\nu$NS\xspace}
\newcommand{\tonyr}{ton$\cdot$yr\xspace}

\preprint{IFT-UAM/CSIC-26-15}

\title{\boldmath Neutrino NSI in archaeological Pb}

\emailAdd{res-nova@unimib.it}

\author[a,b]{D.~Alloni}
\author[c,d]{G.~Benato}
\author[e,f]{P.~Carniti}
\author[e,f]{M.~Cataldo}
\author[q]{D.~Cerde\~{n}o}
\emailAdd{davidg.cerdeno@ift.csic.es}
\author[r,s]{A.~Cheek}
\emailAdd{acheek@sjtu.edu.cn}
\author[g]{L.~Chen}
\author[e,f]{M.~Clemenza}
\author[e,f]{M.~Consonni}
\author[e,f]{G.~Croci}
\author[h]{I.~Dafinei}
\author[c,i,j]{F.A.~Danevich}
\author[b]{C.~de~Vecchi}
\author[e,f]{D.~Di~Martino}
\author[f,k]{E.~Di~Stefano}
\author[e,f]{N.~Ferreiro Iachellini}
\emailAdd{nahuel.ferreiroiachellini@unimib.it}
\author[c,h]{F.~Ferroni}
\author[e,f,k]{F.~Filippini}
\author[q]{P.~Foldenauer}
\emailAdd{patrick.foldenauer@csic.es}
\author[l]{S.~Ghislandi}
\author[e,f]{A.~Giachero}
\author[e,f]{L.~Gironi}
\author[f]{C.~Gotti}
\author[d]{P.~Gorla}
\author[d]{D.L.~Helis}
\author[i]{D.V.~Kasperovych}
\author[i]{V.V.~Kobychev}
\author[e,f]{G.~Marcucci}
\author[d,m]{A.~Melchiorre}
\author[b,n]{A.~Menegolli}
\author[d]{S.~Nisi}
\author[b,o]{M.~Musa}
\author[c,d]{L.~Pagnanini}
\author[e,f]{L.~Pattavina}
\author[f]{G.~Pessina}
\author[d]{S.~Pirro}
\author[f]{S.~Pozzi}
\author[b]{M.C.~Prata}
\author[d]{A.~Puiu}
\author[e,f]{S.~Quitadamo}
\author[b,o]{M.P.~Riccardi}
\author[b]{M.~Rossella}
\author[b,n]{R.~Rossini}
\author[f,p]{E.~Sala}
\author[f,o]{F.~Saliu}
\author[a,b]{A.~Salvini}
\author[d,i,j]{V.I.~Tretyak}
\author[e,f]{L.~Trombetta}
\author[e,f]{D.~Trotta}
\author[g]{H.~Yuan}

\affiliation[a]{Laboratorio Energia Nucleare Applicata,
Via Aselli 41, I-27100 Pavia, Italy}

\affiliation[b]{INFN Sezione di Pavia,
Via Bassi 6, I-27100 Pavia, Italy}

\affiliation[c]{Gran Sasso Science Institute,
Viale F. Crispi 7, I-67100 L’Aquila, Italy}

\affiliation[d]{INFN Laboratori Nazionali del Gran Sasso,
Via G. Acitelli 22, I-67100 Assergi, Italy}

\affiliation[e]{Dipartimento di Fisica, Universit\`a di Milano - Bicocca,
Piazza della Scienza 3, I-20126 Milano, Italy}

\affiliation[f]{INFN Sezione di Milano - Bicocca,
Piazza della Scienza 3, I-20126 Milano, Italy}

\affiliation[g]{Shanghai Institute of Ceramics, CAS,
1295 Dingxi Road, Shanghai 200050, P.R. China}

\affiliation[h]{INFN Sezione di Roma-1,
P.le Aldo Moro 2, I-00185 Roma, Italy}

\affiliation[i]{Institute for Nuclear Research of NASU,
03028 Kyiv, Ukraine}

\affiliation[j]{Institute of Experimental and Applied Physics,
Czech Technical University in Prague,
Husova 240/5, 110 00 Prague 1, Czech Republic}

\affiliation[k]{DISAT, Universit\`a di Milano - Bicocca,
Piazza della Scienza 1, I-20126 Milano, Italy}

\affiliation[l]{Massachusetts Institute of Technology,
Cambridge, MA 02139, USA}

\affiliation[m]{Dipartimento di Scienze Fisiche e Chimiche,
Universit\`a degli Studi dell’Aquila,
I-67100 L’Aquila, Italy}

\affiliation[n]{Dipartimento di Fisica, Universit\`a di Pavia,
Via Bassi 6, I-27100 Pavia, Italy}

\affiliation[o]{Dipartimento di Scienze della Terra e dell'Ambiente,
Universit\`a di Pavia,
Via Ferrata 7, I-27100 Pavia, Italy}

\affiliation[p]{Center for Underground Physics,
Institute for Basic Science,
34126 Daejeon, Korea}

\affiliation[q]{Instituto de F\'isica Te\'orica IFT-UAM/CSIC, 
Cantoblanco, E-28049, Madrid, Spain}

\affiliation[r]{State Key Laboratory of Dark Matter Physics, Tsung-Dao Lee Institute \& School of Physics and Astronomy, Shanghai Jiao Tong University, Shanghai 200240, China}

\affiliation[s]{Key Laboratory for Particle Astrophysics and Cosmology (MOE) \& Shanghai Key Laboratory for Particle Physics and Cosmology, Shanghai Jiao Tong University, Shanghai 200240, China}

\abstract{
Dark matter direct detection experiments can observe solar neutrinos via coherent elastic neutrino-nucleus scattering, making it possible to test new physics in the neutrino sector. In this article, we study the sensitivity of \RESNOVA{}, a novel cryogenic calorimetric experiment employing PbWO$_4$ crystals grown from archaeological lead, to  neutrino non-standard interactions (NSI). 
We perform a sensitivity study for a benchmark setup with a nominal energy threshold of 1~keV and an exposure of 1~\tonyr, both for a conservative (only heat readout) and ideal (heat and scintillation) background rejection scenario.
We find that, in its nominal configuration, while not being sensitive to Standard Model solar $\nu$ interactions, \RESNOVA{} can reach sensitivities to NSI at the level of current global fits.
With moderate or significant improvements of the threshold down to $0.5$ keV and $0.2$ keV, \RESNOVA{} will be able to achieve sensitivities beyond NSI global fit results, 
testing new areas of the parameter space in the electron and tau sectors, $\varepsilon_{ee}$, $\varepsilon_{\tau\tau}$, and $\varepsilon_{e\tau}$. A similar improvement in sensitivities is expected when instead increasing the exposure to 10 \tonyr.
}

\begin{document}
\maketitle
\flushbottom

\section{Introduction}
\label{sec:intro}

The Sun constitutes one of the most intense and well-characterized sources of low-energy neutrinos~\cite{Vitagliano:2019yzm, solarnuAgostini, Bellini:2011rx, Borexino:2007kvk, PhysRevD.108.102005}. Their detection has played a pivotal role in the development of the physics of the electroweak sector~\cite{Bahcall2004}, providing direct insight into solar fusion processes~\cite{Bahcall:2005va} and leading to the discovery of neutrino flavor oscillations~\cite{solar-analysis}. In recent decades, solar neutrinos have been observed using a variety of experimental techniques, including radiochemical detectors~\cite{Davis1968,GALLEX:1998kcz,GNO:2005bds}, water Cherenkov experiments~\cite{Kamiokande1989,Super-Kamiokande:2016yck,SNO2002}, and liquid scintillator/ detectors~\cite{KamLANDSolar2015,PhysRevD.108.102005}. Despite these achievements, several open questions remain, such as the precise determination of CNO-cycle neutrino fluxes~\cite{Haxton2013,Vinyoles:2016djt,PhysRevD.108.102005} and the possible presence of non-standard neutrino interactions (NSI)~\cite{Miranda:2015dra,Gonzalez-Garcia:2016gpq, Farzan:2017xzy,Coloma:2017egw,Esteban:2018ppq,Esteban:2019lfo}, especially in the context of the degeneracies in determining the neutrino mass ordering associated with the LMA-Dark solution~\cite{Denton:2022nol}.

In this context, coherent elastic neutrino--nucleus scattering (CE$\nu$NS) has emerged as a promising detection channel~\cite{Freedman:1973yd,Drukier:1983gj,Cerdeno:2016sfi,Bento:2024zkd}. As a neutral-current process, CE$\nu$NS is sensitive to all active neutrino flavors, providing a flavor-independent probe of solar neutrino fluxes~\cite{Solar_Xe,Solar_Panda} and physics beyond the Standard Model (SM)~\cite{Pandey:2023arh}. CE$\nu$NS detection is particularly challenging for MeV-scale neutrinos from continuous sources. In this regime, the interaction produces only a single, low-energy nuclear recoil (typically at the keV scale), with no coincident signature, making it intrinsically difficult to distinguish from backgrounds. This places stringent requirements on both ultra-low energy thresholds and background suppression~\cite{Baxter:2019mcx}.

Recent advances in cryogenic detector technologies have opened new opportunities for CE$\nu$NS measurements with solar neutrinos~\cite{Essig:2022dfa}. Several next-generation dark matter experiments are now approaching the so-called neutrino fog~\cite{Billard:2013qya,OHare:2021utq}, where CE$\nu$NS from solar neutrinos becomes an irreducible background. In this regime, xenon-based direct dark matter experiments have recently reported the observation of solar neutrinos from the $^8$B flux, among the highest-energy components of the solar spectrum. With increasing statistical significance, PandaX-4T~\cite{Solar_Panda}, XENONnT~\cite{XENON:2024hup}, and LZ~\cite{LZ:2025igz} have employed these data to set new constraints on dark matter properties.

Beyond their role as backgrounds, such measurements can enable precision studies and provide novel tests of fundamental physics in the neutrino sector~\cite{Billard:2014yka,Newstead:2018muu,Dutta:2019oaj}. In particular, they offer a unique opportunity for the direct observation of the solar $\nu_\tau$ flux~\cite{Dutta:2020che}. In this regard, direct detection experiments provide complementary information to flavor-dependent non-standard interactions (NSI), owing to their simultaneous sensitivity to flavor-dependent propagation and scattering effects~\cite{Gonzalez-Garcia:2015qrr,Gonzalez-Garcia:2018dep,Proceedings:2019qno,Amaral:2023tbs,AristizabalSierra:2024nwf,Celestino-Ramirez:2025snn}.

Low-threshold bolometers~\cite{Pirro:2017ecr} can provide a complementary test of solar neutrinos through CE$\nu$NS, benefiting from their low detection thresholds and different choices of target materials.
\RESNOVA{} is a cryogenic observatory designed to exploit CE$\nu$NS for flavor-independent neutrino detection using ultra-low-background archaeological PbWO$_4$ detectors~\cite{Pattavina:2020cqc}. The use of
heavy target nuclei enhances the CE$\nu$NS cross section ($\propto N^2$). However, this comes at the
cost of extremely small nuclear recoil energies, which scale inversely with the nuclear mass,
thereby increasing the demands on detector threshold and background control~\cite{RES-NOVA:2021gqp}.
In this work, we assess the potential of the \RESNOVA{} detector concept to detect solar neutrinos via CE$\nu$NS and probe NSI. While standard solar neutrino fluxes lie close to the experimental sensitivity frontier, \RESNOVA{} remains sensitive to deviations induced by NSI under realistic detector configurations. We investigate the accessible parameter space and discuss the technological requirements necessary to enable precision solar neutrino measurements with cryogenic detectors.

\bigskip

This article is structured as follows. In~\cref{sec:resnova}, we describe the \RESNOVA{} detector design and details of operation. We review the relevant ingredients to compute the solar neutrino scattering signal in PbWO$_4$ in~\cref{sec:theory}. Next, we comment on our statistical limit setting procedure in~\cref{sec:results} and determine our sensitivity projections to NSI from solar neutrino scattering in PbWO$_4$. Finally, we present our conclusions in~\cref{sec:summary}.

\section{The \RESNOVA{} detector concept}
\label{sec:resnova}

\RESNOVA{} is a cryogenic observatory designed to search for neutrinos and dark matter using an array of ultra-low-background detectors based on archaeological Pb~\cite{Pattavina:2020cqc}. Its core technology consists of scintillating PbWO$_4$ crystals operated as cryogenic bolometers at millikelvin temperatures, each instrumented with independent thermal and scintillation light readout channels~\cite{Beeman:2012wz,kg-scale}.

The use of archaeological Pb is motivated by its exceptionally low intrinsic radioactivity, particularly with respect to long-lived isotopes such as $^{210}$Pb, whose activity is strongly suppressed due to centuries of natural decay~\cite{Pattavina:2019pxw}. This property ensures a low background environment suitable for rare-event searches. In addition, the high density of PbWO$_4$ allows for large target masses within compact detector volumes.

\RESNOVA{} is optimized for CE$\nu$NS detection, exploiting the high neutron number of Pb nuclei and the ultra-low energy thresholds achievable with cryogenic sensors~\cite{FerreiroIachellini:2021qgu}. The same experimental infrastructure enables searches for neutrinos from core-collapse supernovae~\cite{RES-NOVA:2021gqp}, solar neutrinos, dark matter~\cite{RES-NOVACollaboration:2025stq}, and axions~\cite{axion}. The modular detector design provides a natural path toward scalability from kilogram-scale demonstrators to multi-ton experiments.

\subsection{Principle of operation}

\RESNOVA{} employs cryogenic calorimetric detection techniques, which allow the measurement of extremely small energy deposits with high sensitivity. Each detector module consists of a PbWO$_4$ crystal absorber operated at mK temperatures, with a heat capacity at the pJ/K scale~\cite{PWO_heat}. Denoting with $C$ such heat capacity, an energy deposition $\Delta E$ produces a temperature rise $\Delta T \sim \Delta E / C$, which is read by a Transition Edge Sensor (TES)~\cite{Pirro:2017ecr}. At cryogenic temperatures, the Debye $T^3$ dependence of $C$ results in a higher signal to the low-energy nuclear recoil characteristic of CE$\nu$NS. The sensitivity of the full system, crystal absorber + TES, to nuclear recoils follows a well-established scaling law~\cite{NUCLEUS:2017gvo} that allows to reach energy thresholds below 1~keV for kg-sized detector modules.

The experiment will be hosted at the underground Gran Sasso National Laboratory (LNGS), benefiting from an overburden of approximately 3800~m.w.e.~\cite{G.Bellini_2012}. \RESNOVA{} has demonstrated 150~eV thresholds using small-mass ($\sim$15\,g) PbWO$_4$ absorbers instrumented with TES sensors~\cite{FerreiroIachellini:2021qgu} and operated at noise conditions worse by a factor 10-100 with respect to well-shielded set-ups~\cite{CRESST:2017ues}, and high count-rate. Despite this, the dector was stably operated for over two days. This threshold value is defined as 5$\times \sigma$, with $\sigma$ being the energy resolution at smallest signals. Based on this proof of principle, a demonstrator with kilogram-scale absorbers and a target threshold of $\sim$1\,keV is under development~\cite{kg-scale}.  Its performance is supported by established scaling laws~\cite{Proebst} and previous results in cryogenic calorimetry~\cite{Strauss:2017cam}.

\begin{figure}
    \centering
    \includegraphics[width=.65\textwidth]{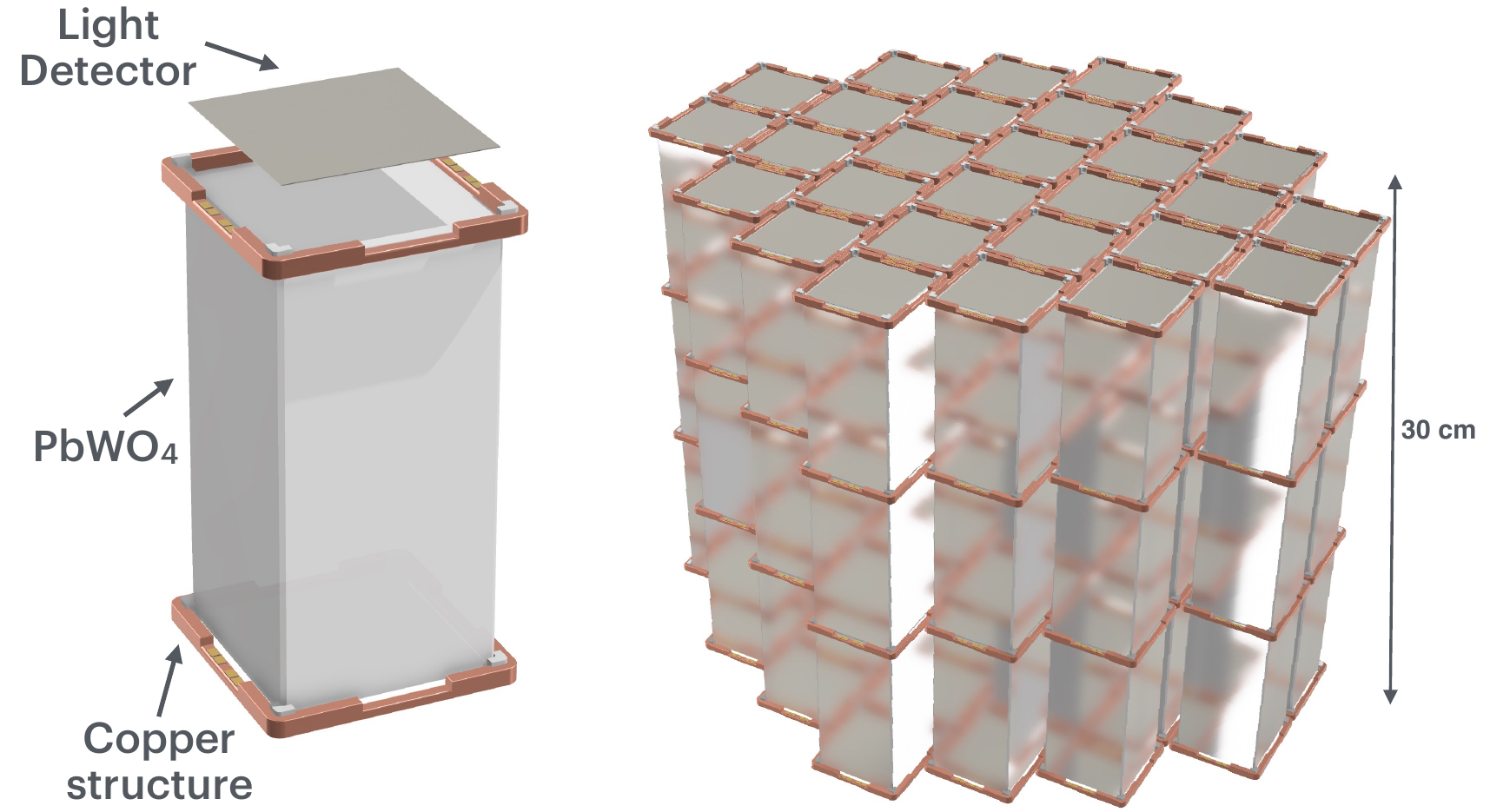}
    \caption{Schematic view of a single detector unit (left) and the full \RESNOVA{} detector (right). The setup includes a PbWO$_4$ crystal absorber, a cryogenic light detector and copper mechanical structure. Individual absorber are foreseen to weigh around 1~kg, while the total mass of the detector's array is expected to fall in the range [170 - 200]~kg.\label{fig:module}}
\end{figure}

Each module integrates a dual readout system consisting of a PbWO$_4$ absorber and a cryogenic light detector made of a thin Ge wafer. The simultaneous measurement of heat and scintillation light enables particle identification on an event-by-event basis. A schematic view of a detector module and the full array is shown in~\cref{fig:module}.

\subsection{Archaeological PbWO$_4$ as cryogenic detector}

The detection of CE$\nu$NS requires ultra-low background levels due to the small interaction cross-sections, typically $10^{-39}$--$10^{-41}~\mathrm{cm}^2$~\cite{Cadeddu:2023tkp}. Backgrounds are dominated by intrinsic radioactivity in detector materials, including isotopes from the $^{238}$U and $^{232}$Th decay chains and $^{210}$Pb~\cite{Laubenstein:2020rbe, Clemenza:2011zz}. Archaeological Pb provides a significant advantage by strongly suppressing cosmogenic contributions, in particular $^{210}$Pb and its progeny~\cite{Pattavina:2019pxw}.

A detailed background prediction for the \RESNOVA{} demonstrator was presented in Ref.~\cite{RES-NOVACollaboration:2025stq}. The model includes contributions from environmental radiation, detector shielding materials, bulk contaminations of detector components, and surface contaminations of elements facing the detector active components. Naturally occurring radioactive isotopes from the $^{232}$Th and $^{238}$U decay chains and the $^{210}$Pb sub-chain are considered, as well as $\gamma$ and neutron environmental backgrounds. As pointed out in Ref.~\cite{RES-NOVACollaboration:2025stq}, bulk crystal contaminations contribute the most to the overall background budget, despite the use of archaeological Pb.

The high mass density of PbWO$_4$, $\rho \simeq 8.3~\mathrm{g\,cm^{-3}}$, enables large target masses in compact volumes, allowing the experimental exposure to be scaled to the $\mathcal{O}(\mathrm{ton}\cdot\mathrm{yr})$ regime while keeping the detector's fiducial volume to the sub-m$^3$ scale. Cryogenic calorimetry also offers flexibility in detector module size, enabling a trade-off between energy threshold, target mass, and channel count. For \RESNOVA{}, modules with masses of $\mathcal{O}(1~\mathrm{kg})$ and thresholds of $\sim$1\,keV are foreseen. By increasing detector granularity, lower thresholds can be achieved while preserving total exposure~\cite{Pirro:2017ecr}, a feature well suited to low-energy nuclear recoil searches.

Two different configurations are considered: a phonon-only readout without particle identification, and an idealized configuration with full $e^-/\gamma$ rejection through combined phonon and scintillation light readout. These scenarios define conservative and ideal benchmarks for the achievable background level, and will be denoted as pessimistic and optimistic scenarios in the following.

\section{CE$\nu$NS with Solar Neutrinos}
\label{sec:theory}

Solar neutrinos are primarily produced in the proton–proton (pp) chain and, at a subdominant level, in the CNO cycle, and they span an energy range from sub-MeV up to several MeV, with the highest-energy component originating from the 
$^8$B decay. Their detection has played a central role in establishing neutrino flavor conversion in the Sun, resolving the long-standing solar neutrino problem~\cite{SNO:2001kpb} and confirming the mechanism of energy generation in the solar core.

In the following, we focus on solar-neutrino detection via CE$\nu$NS in cryogenic PbWO$_4$ detectors and its impact on searches for non-standard neutrino interactions (NSI).

\subsection{Cryogenic detection via CE$\nu$NS}

The detection of solar neutrinos through CE$\nu$NS opens an entirely new experimental approach to solar neutrino physics. Unlike traditional detection via neutrino-electron scattering, which is mostly sensitive to the electron-flavor component, CE$\nu$NS is a neutral-current process and therefore provides sensitivity to the total active neutrino flux, independent of flavor.

This makes CE$\nu$NS detection particularly valuable for testing the flavor conversion mechanisms predicted by the Mikheyev–Smirnov–Wolfenstein (MSW) effect~\cite{Wolfenstein:1977ue,Mikheyev:1985zog}. CE$\nu$NS offers a unique handle on the detection of CNO neutrinos, which remain poorly constrained but are crucial for resolving the long-standing solar metallicity problem.

Cryogenic detectors based on PbWO$_4$ provide a favorable target due to the high neutron number of Pb and W, which enhances the CE$\nu$NS cross-section. Combined with ultra-low energy thresholds and exceptional radio-purity through the use of archaeological Pb, \RESNOVA{} has the potential to perform high-statistics measurements in a background-limited regime. These conditions are precisely those highlighted in Ref.~\cite{Amaral:2023tbs} as necessary to achieve percent-level precision on low-energy solar neutrino fluxes.

Moreover, such measurements would serve as a benchmark for new-generation CE$\nu$NS experiments and contribute to global efforts to combine solar neutrino data from both charged- and neutral-current interactions. The ability to operate in a modular and scalable cryogenic architecture further enhances \RESNOVA{}’s role in the long-term roadmap of solar neutrino precision measurements.

\subsection{Solar Neutrino rates according to SM predictions}

\begin{figure*}[t]
  \includegraphics[width=.45\textwidth]    {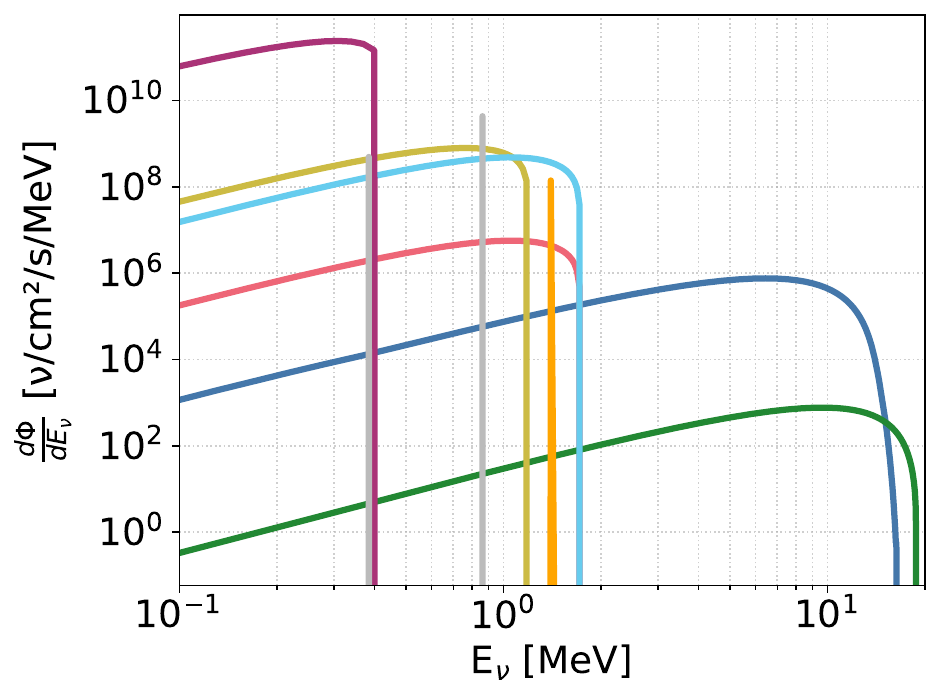} \;\hfill
  \includegraphics[width=.55\textwidth]    {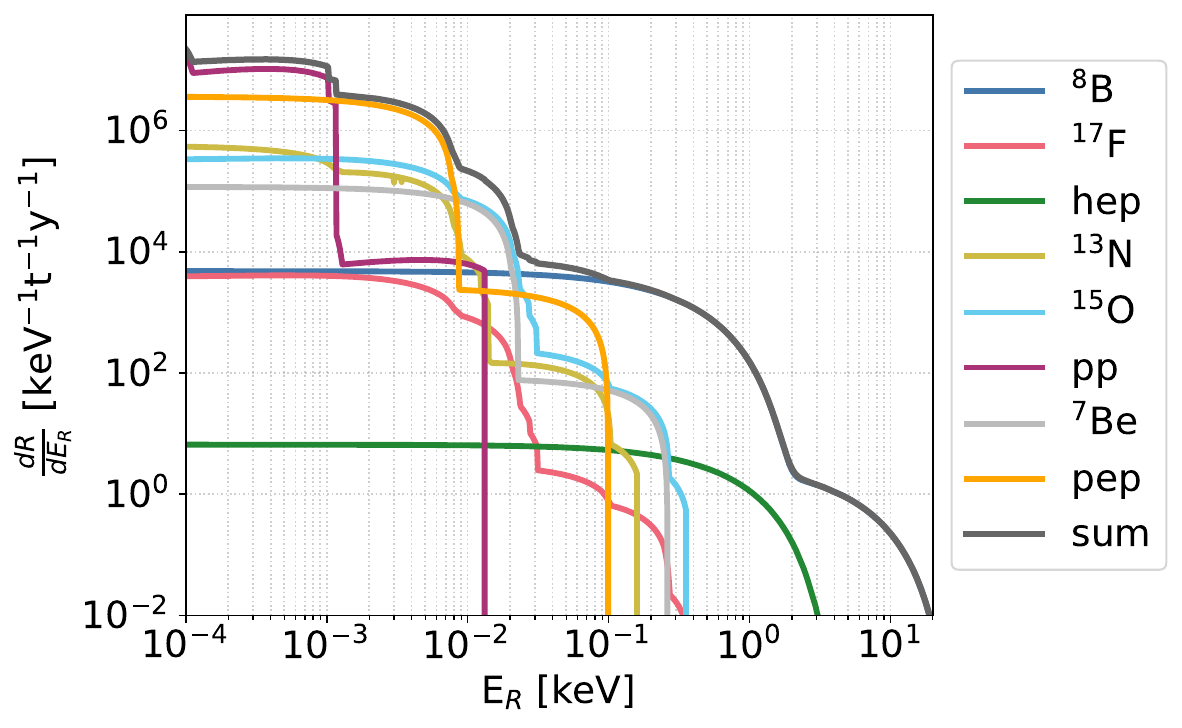}
  \caption{Differential solar neutrino fluxes at earth (left) and corresponding induced recoil spectra in PbWO$_4$ (right). The colored lines correspond to individual flux components (see legend). The black line represents the total expected recoil rate.
  \label{fig:nu_sig}}
\end{figure*}

In general, the differential rate of solar neutrino scattering in a target material $T$ can be expressed as
\begin{align}\label{eq:rate_gen}
    \frac{\mathrm{d} R}{\mathrm{d} E_R}=N_T \int_{E_\nu^{\min }} \frac{\mathrm{d} \Phi_\nu}{\mathrm{d} E_\nu}\, \operatorname{Tr}\left[\boldsymbol{\rho}\, \frac{\mathrm{d} \boldsymbol{\zeta}}{\mathrm{d} E_R}\right] \mathrm{d} E_\nu \,,
\end{align}
where $N_{T}$ is the number of scattering targets, ${\mathrm{d} \Phi_\nu}/{\mathrm{d} E_\nu}$ denotes the differential solar neutrino flux, the components of which we show in the left panel of \cref{fig:nu_sig}~\cite{Bahcall:2004pz}, $\boldsymbol{\rho}$ is the neutrino density matrix encapsulating the flavor mixing and quantum coherence of the incoming solar neutrinos, and ${\mathrm{d} \boldsymbol{\zeta}}/{\mathrm{d} E_R}$ is the generalized neutrino scattering cross section as defined in Refs.~\cite{Coloma:2022umy,Amaral:2023tbs}.

In the SM, neutrino interactions are both flavor-diagonal and flavor-universal, such that the generalized scattering cross section ${\mathrm{d} \boldsymbol{\zeta}}/{\mathrm{d} E_R}$ is diagonal and proportional to the unit matrix. In this case, the neutrino scattering rate greatly simplifies and for our composite target material of PbWO$_4$ reads
\begin{equation}\label{eq:recoils}
\frac{dR}{dE_R} =  \sum_{i,j} N_{Tj} \int_{E_\nu^\text{min}} \frac{\mathrm{d} \Phi_i}{\mathrm{d} E_\nu} \, \frac{d\sigma_j(E_\nu, E_R)}{dE_R}  \, dE_\nu,
\end{equation}
with \( i \in \{\mathrm{pp},\, \mathrm{hep},\, \mathrm{pep},\, {}^7\mathrm{Be},\, {}^{17}\mathrm{F},\, {}^8\mathrm{B},\, {}^{13}\mathrm{N},\, {}^{15}\mathrm{O} \} \) labeling the different solar neutrino flux components,
\( j \in \{ {}^{208}\mathrm{Pb},\, {}^{183}\mathrm{W},\, {}^{16}\mathrm{O} \} \) the different atomic isotopes,
and $N_{Tj}$ the number of targets of species $j$. Note that since all neutrino flavors contribute to the \cevns rate and SM neutrino interactions are flavor-universal, \cref{eq:recoils} does not depend on the neutrino density matrix.

In the SM, the differential \cevns cross section is given in the familiar form as~\cite{Freedman:1973yd}\footnote{Throughout this paper we work in natural units, setting $\hbar=c=1$.}
\begin{eqnarray}
\label{eq:xsec}
\frac{d\sigma}{d E_R} = \frac{G^2_F m_N}{4 \pi}  
\, \left(1- \frac{E_R\, m_{N}}{2E_{\nu}^2} \right) \,Q_{\nu N}^2 \, |F(q)|^2 ,
\end{eqnarray}
where $G_F$ is the Fermi coupling constant,
${Q_{\nu N} = N - (1 -4\,\sin^2\theta_W)\,Z}$ is the SM coherence factor with
$Z$ and $N$ the atomic and neutron numbers of the target nucleus, $m_N$ its mass,  $\theta_W$ the Weinberg angle, $E_{\nu}$ the neutrino energy and $E_R$ the recoil energy of the target. Finally, $F(q)$, is the Helm or elastic nuclear form factor~\cite{Helm:1956zz, Lewin:1995rx} at momentum transfer $q=\sqrt{2E_R m_N}$, also known as \textit{coherence factor}.\\

\begin{figure*}[t]
\centering
\includegraphics[width=.7\textwidth]{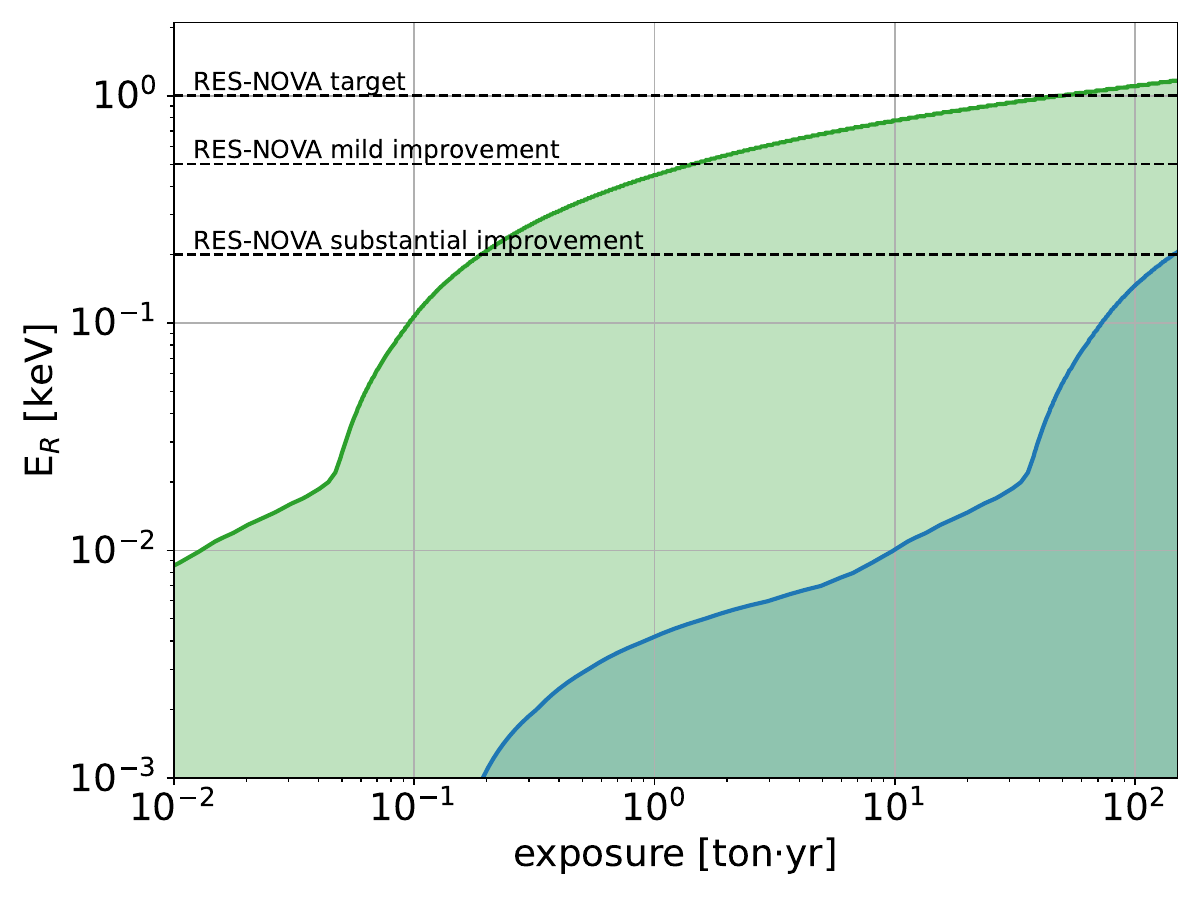}
\caption{The 3$\sigma$ C.L. discovery reach for solar neutrinos via SM CE$\nu$NS, as a function of different energy thresholds and exposures, computed assuming the best (green) and worst (blue) background scenarios for PbWO$_4$ detectors in the \RESNOVA{} installation. The range of energy thresholds considered corresponds to values that are technologically achievable with cryogenic detectors, as discussed in Ref.~\cite{Bento:2024zkd}. Three representative thresholds are highlighted: 1~keV, adopted as the \RESNOVA{} benchmark configuration; and 0.5~keV and 0.2~keV, which can be reached by trading target mass for increased detector granularity and readout channel count. The increase in sensitivity in the pessimistic scenario between thresholds of around 0.1 keV and 0.2 keV is due to additional neutrino flux components entering the signal region (cf.~right panel of~\cref{fig:nu_sig}).}
\label{fig:thr_vs_exp}
\end{figure*}

The resulting nuclear recoil rate for a PbWO$_4$ target is shown in the right panel of~\cref{fig:nu_sig}, where we show the total rate and the individual contributions from different neutrino flux sources.
These calculated rates are in line with those from Ref~\cite{Bento:2024zkd} for CaWO$_4$ or Al$_2$O$_3$ under similar conditions.
Assuming a detector threshold of 1~keV and using standard solar model fluxes~\cite{Bahcall:2004pz}, a 1-tonne \RESNOVA{} detector operated over 1~yr is expected to register 35 solar neutrinos induced events. These are induced mainly by $^8$B  and with only a small contribution from $\it hep$ neutrinos. Such event rate lies at the edge of detectability in the optimistic background scenario (see~\cref{fig:thr_vs_exp}).

\RESNOVA{} represents a promising new direction in solar neutrino detection. The \RESNOVA{} demonstrator will play a critical role in validating threshold performance and background assumptions, paving the way for a scalable, cryogenic, and flavor-independent solar neutrino observatory.

\subsection{Non-Standard Interactions}

We will evaluate the sensitivity of \RESNOVA{} to new neutrino physics in the effective field theory framework on non-standard interactions (NSI), which are characterised by the Lagrangian~\cite{Wolfenstein:1977ue,Guzzo:1991hi,Guzzo:1991cp,Gonzalez-Garcia:1998ryc,Bergmann:2000gp,Guzzo:2000kx,Guzzo:2001mi,Gonzalez-Garcia:2004pka,Gavela:2008ra}
\begin{equation}\label{eq:eff_op}
    \mathcal{L}_\text{NSI} = -2\sqrt{2}\, G_F \sum_{\substack{f=e,u,d\\\alpha,\beta=e,\mu,\tau}} \varepsilon^{fP}_{\alpha\beta} \ \left[\bar\nu_\alpha \gamma_\rho P_L \nu_\beta\right] \, \left[\bar f \gamma^\rho P f \right]\,,
\end{equation}
with $G_F$ denoting the Fermi constant, $P\in\{P_L,P_R\}$ the chirality projectors and $\varepsilon^{fP}_{\alpha\beta}$ the NSI coefficients. Neutrino-matter effects are only present for the vector part of the interactions, $\varepsilon^{f}_{\alpha\beta} = \varepsilon^{fL}_{\alpha\beta} + \varepsilon^{fR}_{\alpha\beta}$. 
Since the solar medium consists only of ordinary matter made up of first-generation fermions, it is sufficient to consider NSI with electrons, up- and down-quarks, or equivalently, electrons, protons and neutrons, $f\in\{e,p,n\}$. This spans a three-dimensional sub-space in the NSI parameter space. Following~\cite{Amaral:2023tbs} and assuming that the NSI neutrino flavor structure is independent of the fermion flavor structure, we  can conveniently  parametrize  a general NSI element by two angles as
\begin{equation}\label{eq:nsi_param}
    \varepsilon^{f}_{\alpha\beta} = \varepsilon^{\eta,\varphi}_{\alpha\beta} \ \xi^f\,,
\end{equation}
with the three base directions defined as
\begin{align}
    \xi^e &= \sqrt{5}\, \cos \eta \, \sin  \varphi \,, \\
    \xi^p &= \sqrt{5}\, \cos \eta \, \cos  \varphi \,, \\
    \xi^n &= \sqrt{5}\, \sin \eta  \,,
\end{align}
with $\eta$ being the angle in the proton-neutron plane, and $\varphi$ the angle in the proton-electron plane.

In the NSI framework, the matter Hamiltonian driving neutrino oscillations in the solar and terrestrial medium gets modified to~\cite{Farzan:2017xzy}
\begin{equation}
 H_\mathrm{mat} = \sqrt{2 } G_F \, N_e(x) \, 
 \left(
 \begin{matrix}
1 + \mathcal{E}_{ee}(x) & \mathcal{E}_{e\mu}(x) & \mathcal{E}_{e\tau}(x) \\
\mathcal{E}_{e\mu}^*(x) & \mathcal{E}_{\mu\mu}(x) & \mathcal{E}_{\mu\tau}(x) \\
\mathcal{E}_{e\tau}^*(x) & \mathcal{E}_{\mu\tau}^*(x) & \mathcal{E}_{\tau\tau}(x) \\
 \end{matrix}
 \right), 
\end{equation}
with the effective in-medium NSI elements
\begin{equation}
    \mathcal{E}_{\alpha\beta} = \sum_{f} \frac{N_f(x)}{N_e(x)} \, \varepsilon^{f}_{\alpha\beta}\,,
\end{equation}
where $N_f(x)$ is the spatial fermion density in matter. 
Since in neutral matter we have $N_p(x)=N_e(x)$, effective in-medium NSI elements can be simply expressed as
\begin{equation}\label{eq:full_nsi}
    \mathcal{E}_{\alpha\beta} = \varepsilon^e_{\alpha\beta} + \varepsilon^p_{\alpha\beta} + Y_n(x)\, \varepsilon^n_{\alpha\beta}  = 
    \left[\xi^e + \xi^p  + Y_n(x)\, \xi^n\right] \, \varepsilon^{\eta,\varphi}_{\alpha\beta} \,,
\end{equation}
where $Y_n(x)={N_n(x)}/{N_e(x)}$ denotes  the relative neutron-to-proton abundance, which we take from Ref.~\cite{Bahcall:2005va}.

Following Ref.~\cite{Amaral:2023tbs}, we can find the solar neutrino density matrix (of initially pure electron-neutrinos) at Earth from
\begin{align}
    \rho^{(e)} = S\, \pi^{(e)}\, S^\dagger = 
    \begin{pmatrix}
        |S_{11}|^2 &  S_{11}\, S_{21}^* & S_{11}\, S_{31}^* \\
        S_{11}^*\, S_{21} &  |S_{21}|^2 & S_{21}\, S_{31}^* \\
        S_{11}^*\, S_{31} &  S_{21}^*\, S_{31} & |S_{31}|^2 
    \end{pmatrix} \,,
\end{align}
from the three $S$-matrix components,\footnote{Note that for the purpose of this study we assume $\delta_{\rm CP}=0$ to make our results directly comparable to the NSI global fits in Refs.~\cite{Coloma:2019mbs,Coloma:2023ixt}.
Furthermore, we assume adiabatic evolution of the neutrino states in the solar medium, analogous to the treatment in Ref.~\cite{Amaral:2023tbs}. }
\begin{align}
    S_{11} & =   e^{-i\,\Phi_{33}}\, s_{13}^2  + c_{13}^2 \, \left( e^{i\,\phi}  \, c_{12} \,  c_{m}  +  e^{-i\,\phi}   \, s_{12} \,  s_{m} \right) \,,\\ 
    S_{21} & = c_{13} \left[   s_{13}\, s_{23}\,(e^{i\,\Phi_{33}} - e^{i\,\phi}\, c_{12} \, c_{m} -  e^{-i \,\phi} \,s_{12} \, s_{m}) + e^{-i\,\phi} \, c_{23}\,( c_{12} \, s_{m} -  e^{ 2 i\,\phi} \,s_{12} \, c_{m} ) \right] \,, \\
    S_{31} & = c_{13} \left[ s_{13}\, c_{23}\,(e^{i\,\Phi_{33}} - e^{i\,\phi}\, c_{12} \, c_{m} -  e^{-i\,\phi} \,s_{12} \, s_{m}) - e^{-i\,\phi} \, s_{23}\,( c_{12} \, s_{m} - e^{i\,2\phi} \,s_{12} \, c_{m} )\right] \,,
\end{align}
where $c_{ij}$ and $s_{ij}$ refer to $\cos \theta_{ij}$ and $\sin\theta_{ij}$, while $c_{m}$ and $s_{m}$ refer to the $\cos \theta_{12}^m$ and $\sin\theta_{12}^m$ of the matter mixing angle $\theta_{12}^m$ in the solar medium, respectively. Here, the two phases $\Phi_{33} =  \frac{\Delta m^2_{31}}{2\, E_\nu} L$ and
$\phi = \int_{0}^L \Delta E_{21}^m(x)\, \dl x $ are related to the propagation of the neutrinos from their production points in the solar core to the solar surface, and are ultimately averaged out due to the uncertainty in the exact production point and corresponding propagation length $L$~\cite{Amaral:2023tbs}.
The sine and cosine of the matter mixing angle can be found as
\begin{equation}
\begin{aligned}
    \sin 2\theta_{12}^m = \frac{p}{\sqrt{p^2+q^2}}\,,  && 
    \cos 2\theta_{12}^m &= \frac{q}{\sqrt{p^2+q^2}}\,,
    \label{eq:mat_angles}
\end{aligned}
\end{equation}
with the quantities
\begin{equation}
\begin{aligned}
    p \equiv&~\sin 2\theta_{12}   + 2\,\varepsilon^{\eta, \varphi}_N \left[\xi^e + \xi^p + Y_n(x)\,\xi^n\right] \, \frac{A_{\mathrm{cc}}}{\Delta m_{21}^2}\,, \\
    q \equiv&~ \cos 2\theta_{12} + \left( 2\,\varepsilon^{\eta, \varphi}_D \left[\xi^e + \xi^p + Y_n(x)\,\xi^n\right] - c_{13}^2\right)\, \frac{A_{\mathrm{cc}}}{\Delta m_{21}^2}\,,
\end{aligned}
\end{equation}
with the matter potential $A_\text{cc}= 2\,E_\nu\, V_\text{cc}= 2\,E_\nu\sqrt{2} G_F N_e(x)$ and the effective NSI parameters 
\begin{equation}
    \begin{aligned}
    \varepsilon_{D}^{\eta, \varphi} \equiv\,& c_{13}\, s_{13} \left(s_{23}\, \varepsilon_{e \mu}^{\eta, \varphi}+c_{23} \,\varepsilon_{e \tau}^{\eta, \varphi}\right)-\left(1+s_{13}^{2}\right) c_{23}\, s_{23}\, \varepsilon_{\mu \tau}^{\eta, \varphi} \\
    &-\frac{c_{13}^{2}}{2}\left(\varepsilon_{e e}^{\eta, \varphi}-\varepsilon_{\mu \mu}^{\eta, \varphi}\right)+\frac{s_{23}^{2}-s_{13}^{2}\, c_{23}^{2}}{2}\left(\varepsilon_{\tau \tau}^{\eta, \varphi}-\varepsilon_{\mu \mu}^{\eta, \varphi}\right)\,, \\
    \varepsilon_{N}^{\eta, \varphi} \equiv\, & c_{13}\left(c_{23} \,\varepsilon_{e \mu}^{\eta, \varphi}-s_{23}\, \varepsilon_{e \tau}^{\eta, \varphi}\right)+s_{13}\left[s_{23}^{2}\, \varepsilon_{\mu \tau}^{\eta, \varphi}-c_{23}^{2}\, \varepsilon_{\mu \tau}^{\eta, \varphi}+c_{23}\, s_{23}\left(\varepsilon_{\tau \tau}^{\eta, \varphi}-\varepsilon_{\mu \mu}^{\eta, \varphi}\right)\right]\,.
    \end{aligned}
\end{equation}

In the presence of NSI, the generalized coherent elastic neutrino-nucleus scattering (\cevns) cross section reads~\cite{Amaral:2023tbs},
\begin{align}\label{eq:sig_CEVNS_gen}
    \left(\diff{\zeta_{\nu N}}{E_R}\right)_{\alpha\beta} 
    & = \frac{G_F^2\, m_N}{\pi}\left(1-\frac{m_N\, E_R}{2 E_\nu^2}\right)\ \left[ \frac{1}{4} \, Q_{\nu N}^2\, \delta_{\alpha\beta} - Q_{\nu N} \,G^\mathrm{NSI}_{\alpha\beta} + \sum_\gamma G^\mathrm{NSI}_{\alpha\gamma}G^{\mathrm{NSI}}_{\gamma\beta}\right]\, F^2(E_R) \,,
\end{align}
with the NSI nuclear coupling given by
\begin{align}\label{eq:g_cevns}
G^\mathrm{NSI}_{\alpha\beta} \equiv \left(2\,\varepsilon^{u}_{\alpha\beta} + \varepsilon^{d}_{\alpha\beta} \right) Z +  \left(\varepsilon^{u}_{\alpha\beta} + 2\,\varepsilon^{d}_{\alpha\beta} \right) N 
= (\xi^p\, Z + \xi^n \, N)\ \varepsilon_{\alpha\beta}^{\eta,\varphi} \,.
\end{align}
Note that the contribution from Beyond the SM (BSM) can destructively interfere with the SM one. 
This can lead to \textit{blind spots} in the NSI parameter space at non-zero NSI couplings\footnote{See also the detailed discussion in Sec.~IV.2 of Ref.~\cite{Amaral:2023tbs} and a previous discussion in Ref.~\cite{Dutta:2020che}.}, as can be seen by the  gaps in the sensitivity bands in~\cref{fig:nsisensitivity_0.5kev_1ton,fig:nsisensitivity_0.1kev_1ton,fig:nsisensitivity_1kev_1ton}.

\section{Sensitivity projections}
\label{sec:results}

In this section, we will detail our statistical treatment for deriving limits on the NSI parameters from \cevns and present our resulting sensitivity projections for \RESNOVA{}.

\subsection{Statistical treatment}

To evaluate the RES-NOVA sensitivity to NSIs, we adopt a frequentist approach based on the profile likelihood ratio. We vary only one NSI parameter at a given time to exhibit \RESNOVA{}'s potential in specific directions of the parameter space. The parameters that we scan over are $(\eta,\epsilon)$, while $\varphi$ is set to 0, since we do not expect a significant contribution in electron recoils. Sensitivity projections are obtained using an Asimov data set corresponding to the SM expectation as well as the central values from solar neutrino fluxes according to standard solar models~\cite{Vinyoles:2016djt}.

We treat the data as $i$ discrete energy bins, which will contain some predicted signal, determined by 
\begin{equation}
    S_i^\nu(\eta,\epsilon) = \mathcal{E}\int_{E_1^i}^{E_2^i} \frac{d R}{dE_R}(\eta,\epsilon)\,dE_R,
\end{equation}
the differential recoil rate is given by \cref{eq:rate_gen} and $\mathcal{E}$ is the exposure of the experiment. In addition the background events, $B_i$, give the total predicted events
\begin{equation}
\nu_i(\eta,\epsilon,\delta) = B_i + (1+\delta)\,S^{\nu}_i(\eta,\epsilon) \, ,
\label{eq:nu_i}
\end{equation}
here we have explicitly included the NSI parameters $\eta$ and $\epsilon$. We have also included $\delta$ as a pull parameter to account for a systematic uncertainty on the expected signal. This is primarily lead by 8.5\%~\cite{Bahcall:2004pz} in the $^8$B flux that we conservatively round up to 10\%. To obtain the Asimov dataset we take the central neutrino flux values and the SM expectation, $n_i=\nu_i(\eta=0, \,\epsilon=0,\, \delta=0 )$.

Assuming Poisson statistics in each bin, the likelihood function is defined as
\begin{equation}
\mathcal{L}(\eta,\epsilon,\delta)=
\left[\prod_i \mathrm{Pois}\!\left(n_i \mid \nu_i(\eta,\epsilon,\delta)\right)\right]
\times
\exp\!\left(-\frac{\delta^2}{2\sigma_\delta^2}\right)\, .
\label{eq:likelihood}
\end{equation}
For each $(\eta,\epsilon)$ point, the likelihood is profiled with respect to $\delta$, yielding the conditional maximum-likelihood estimator $\hat{\hat{\delta}}(\eta,\epsilon)$. The profile-likelihood-ratio test statistic is then constructed as
\begin{equation}
q(\eta,\epsilon)=
-2\ln\frac{\mathcal{L}\!\left(\eta,\epsilon,\hat{\hat{\delta}}(\eta,\epsilon)\right)}
{\mathcal{L}\!\left(\hat{\eta},\hat{\epsilon},\hat{\delta}\right)}\, ,
\label{eq:q_etaeps}
\end{equation}
where $(\hat{\eta},\hat{\epsilon},\hat{\delta})$ denote the values that globally maximize the likelihood on the scanned grid.

The solar neutrino signal expectations are computed using the \texttt{SNuDD} package~\cite{snudd2023}, which accounts for the detector response in terms of energy resolution and efficiency. Under Wilks' theorem, $q(\eta,\epsilon)$ asymptotically follows a $\chi^2$ distribution with two degrees of freedom, since $(\eta,\epsilon)$ are treated as the parameters of interest, we draw the $90\%$ confidence level (C.L.) contours in the $(\eta,\epsilon)_{\varphi=0}$ plane using the threshold, which corresponds to $q(\eta,\epsilon)=\chi^2_{2,\,0.90}=4.61$. We have also tested the validity of the Wilk's theorem against $\simeq$400 toy montecarlo simulations for each configuration considered in the following.

\subsection{Analysis results}

\begin{figure}[htbp]
\centering
\includegraphics[width=.49\textwidth]{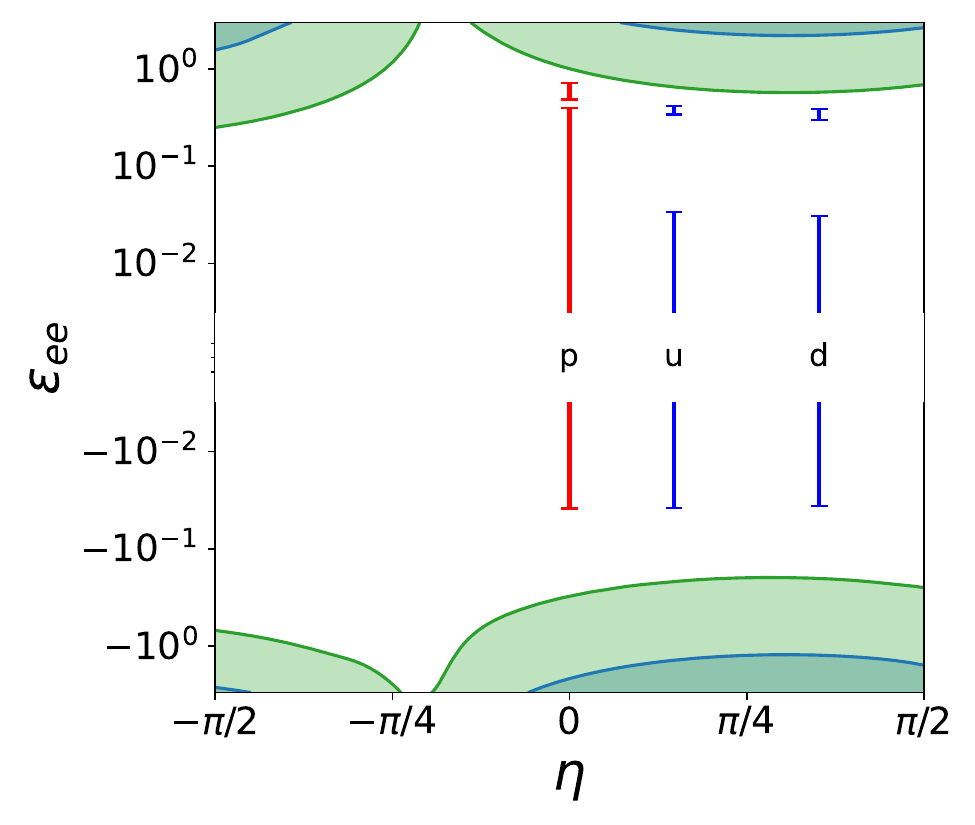}
\includegraphics[width=.49\textwidth]{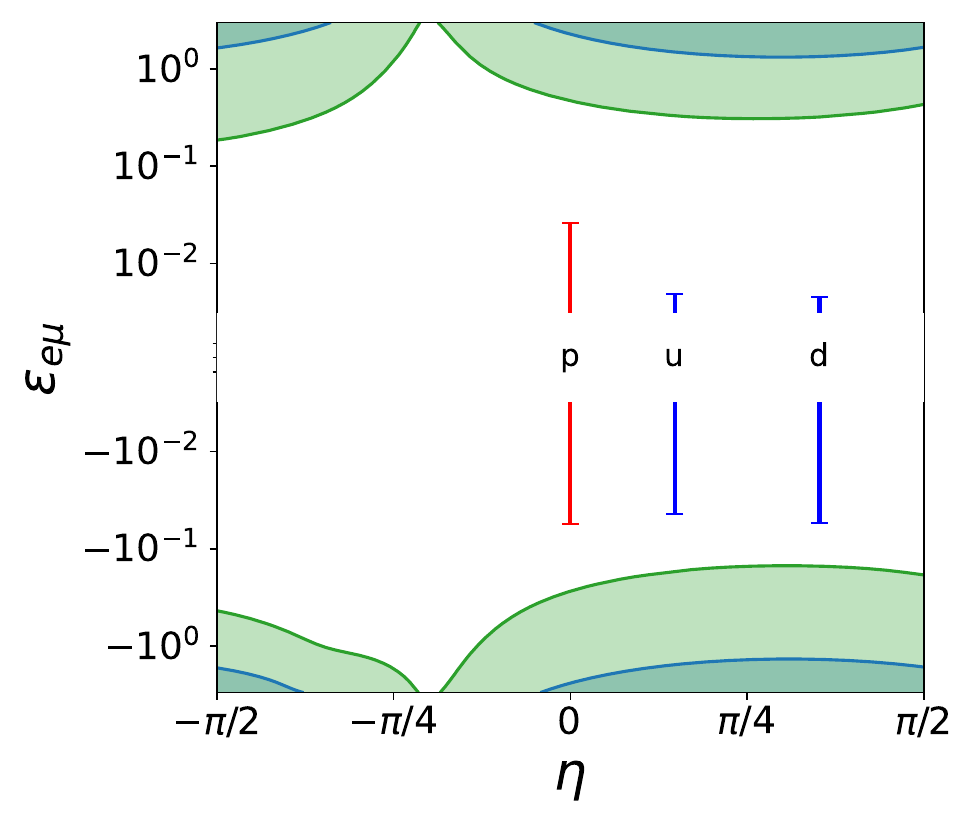} \\
\includegraphics[width=.49\textwidth]{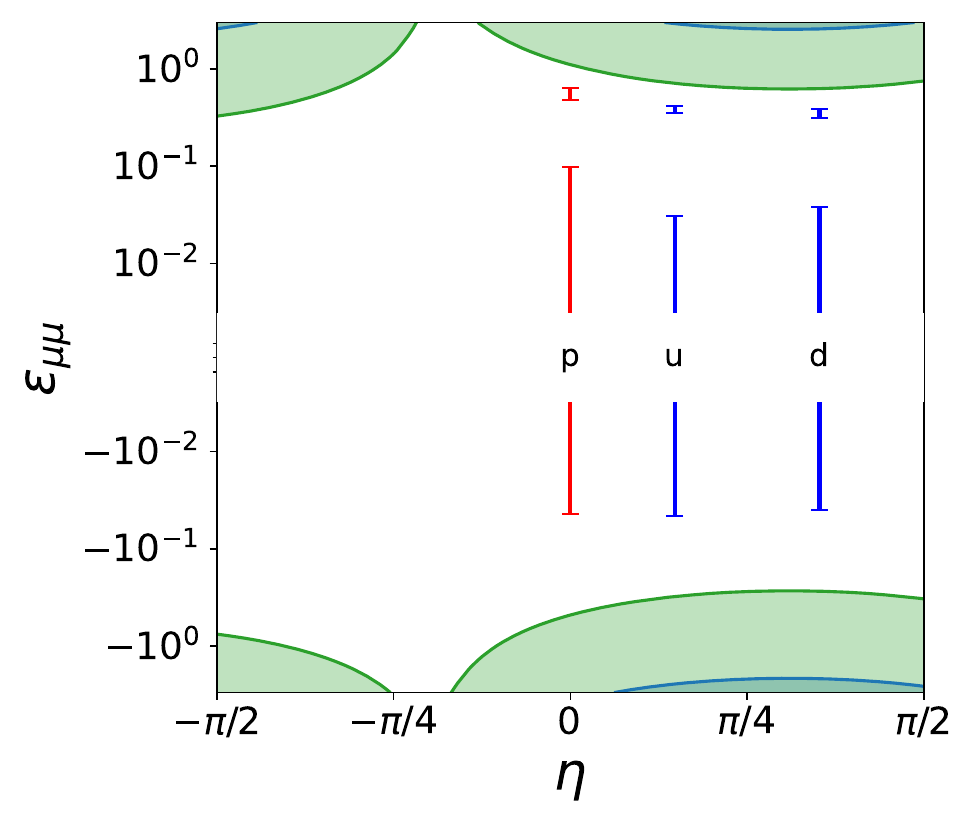}
\includegraphics[width=.49\textwidth]{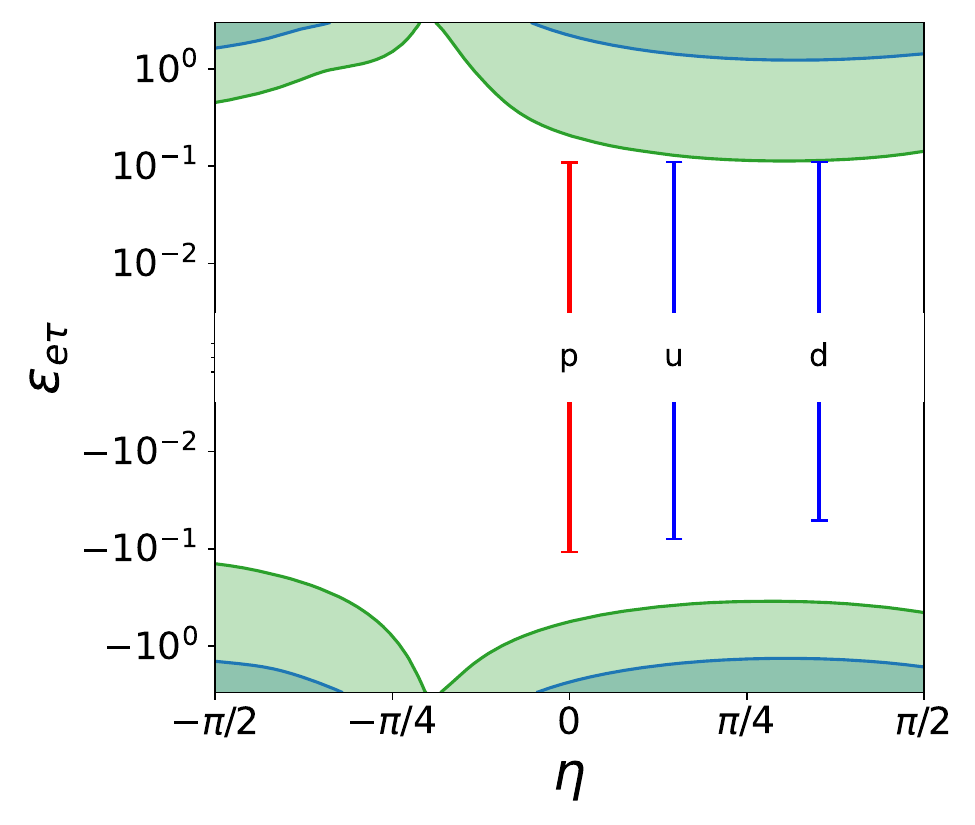} \\
\includegraphics[width=.49\textwidth]{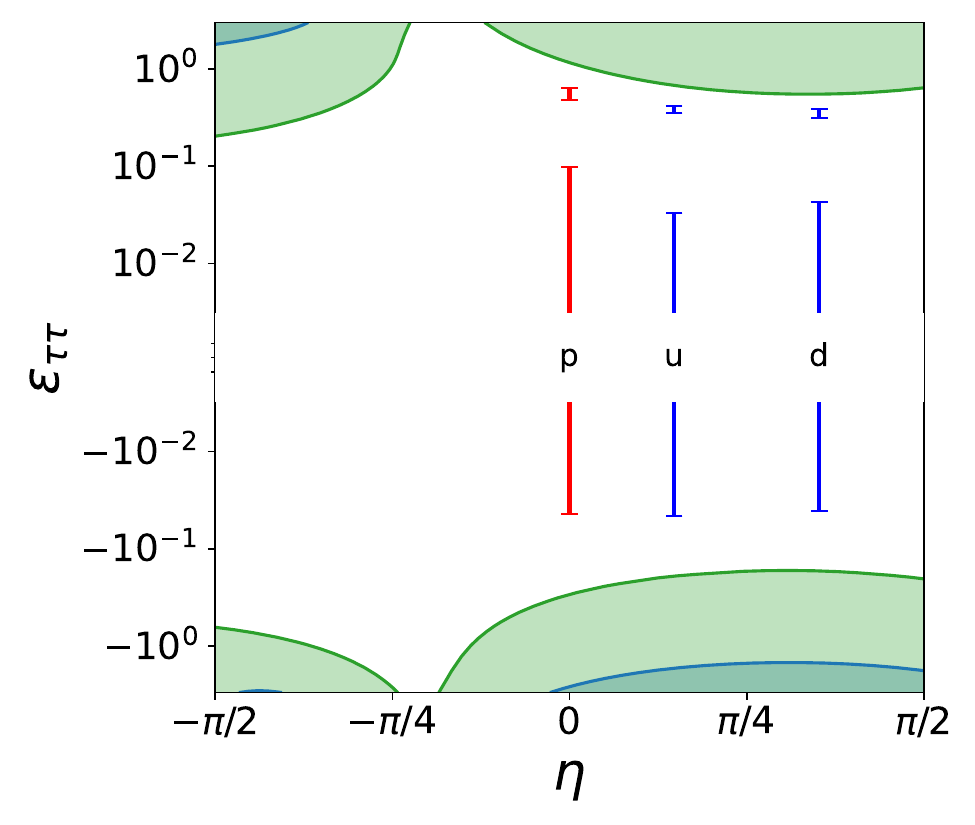}
\includegraphics[width=.49\textwidth]{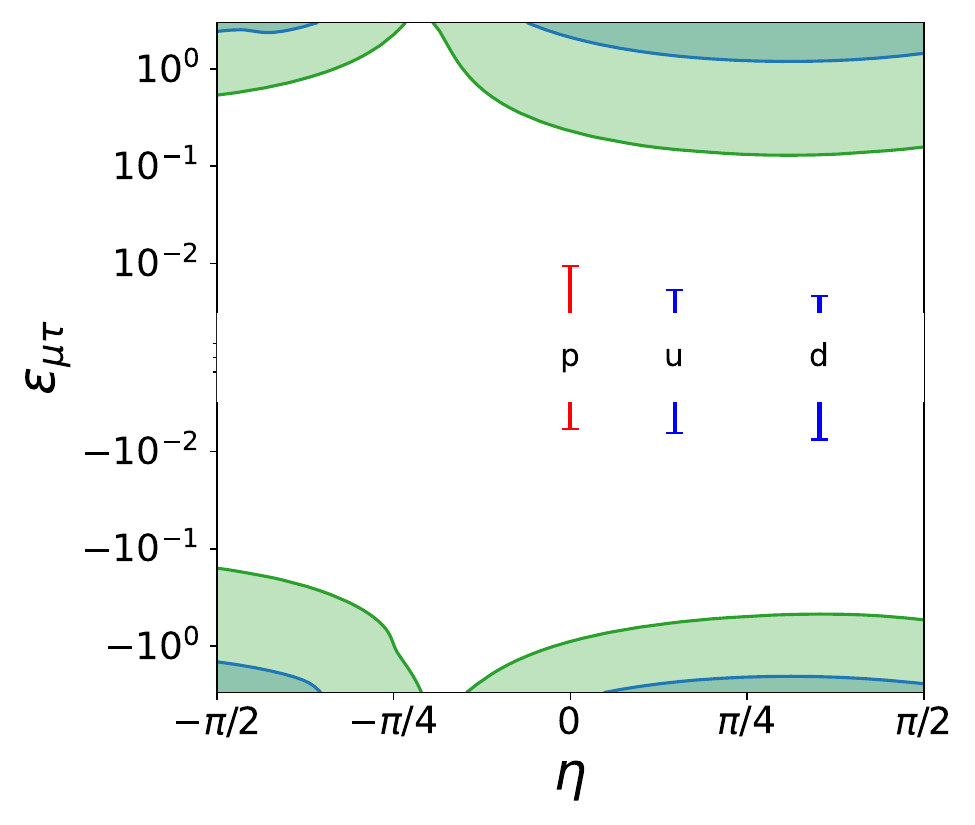}
\caption{The 90\% C.L.~sensitivity of the \RESNOVA{} experiment to NSI with nucleons ($\varphi = 0$) due to nuclear recoils induced via \cevns.
Shown are the projected constraints for two different representative background configurations: an optimistic scenario (green), corresponding to full e-$\gamma$ background rejection, and a pessimistic scenario (blue), based on no background rejection capability. Both projections are obtained for the same detector exposure and energy threshold of 1~\tonyr and 1~keV, respectively. The two configurations illustrate the impact of background control on the sensitivity to NSI in PbWO$_4$ cryogenic detectors. For comparison, we show the $2\sigma$ credible intervals from the NSI global fits in the proton direction, $\varepsilon^p_{\alpha\beta}$, from Ref.~\cite{Coloma:2019mbs} (red), and on the up- and down-quark directions, $\varepsilon^u_{\alpha\beta}$ and $\varepsilon^d_{\alpha\beta}$, from Ref.~\cite{Coloma:2023ixt} (blue).
}
\label{fig:nsisensitivity_1kev_1ton}
\end{figure}

\begin{figure}[htbp]
\centering
\includegraphics[width=.49\textwidth]{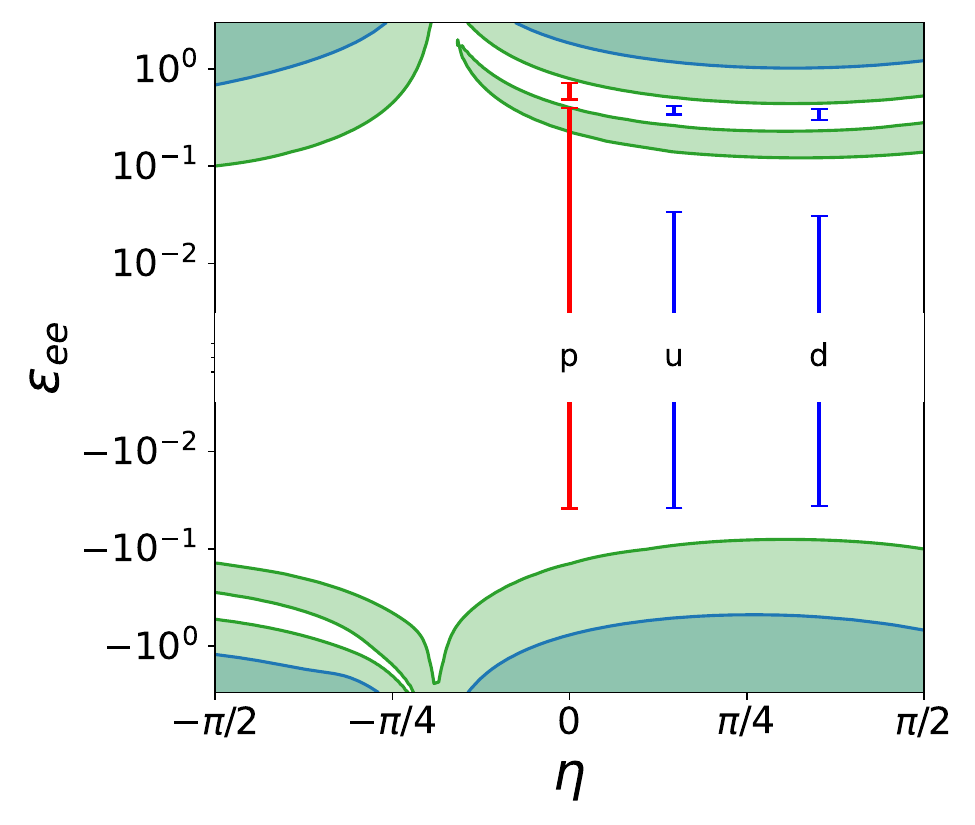} 
\includegraphics[width=.49\textwidth]{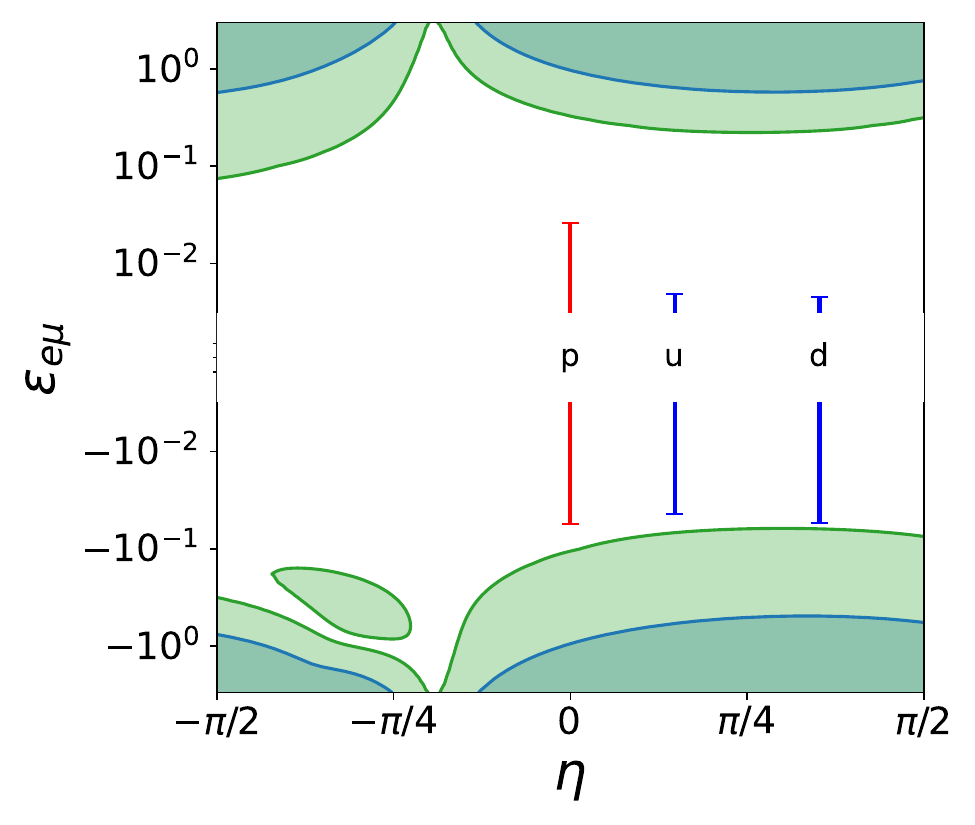} \\
\includegraphics[width=.49\textwidth]{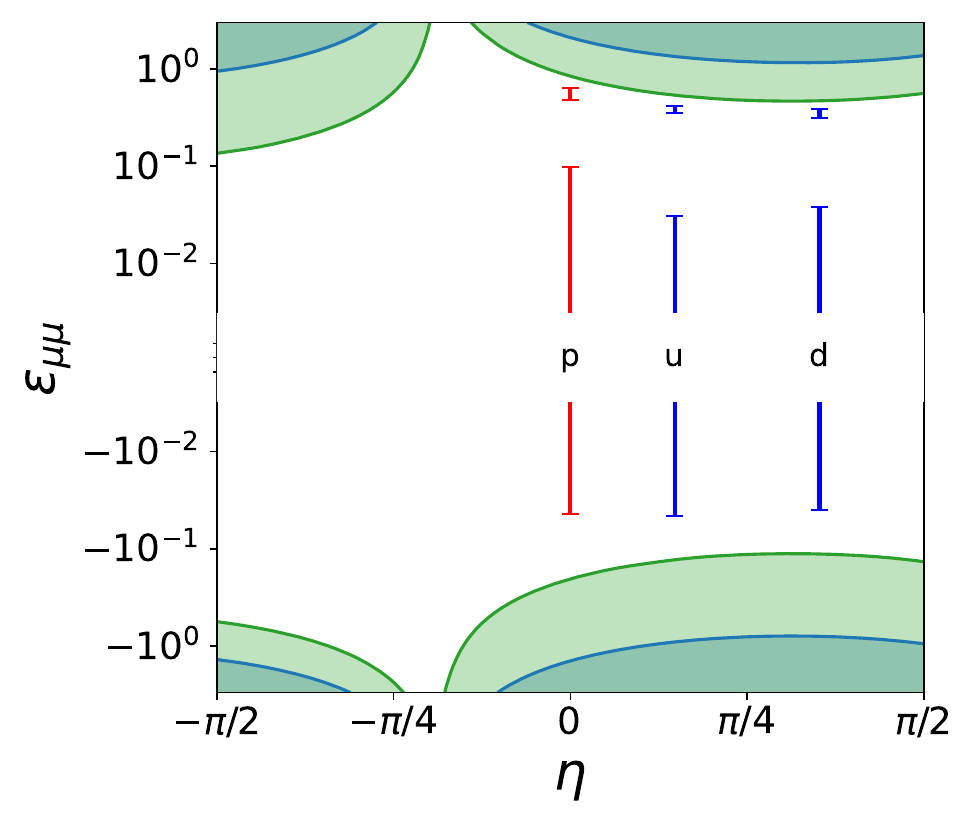} 
\includegraphics[width=.49\textwidth]{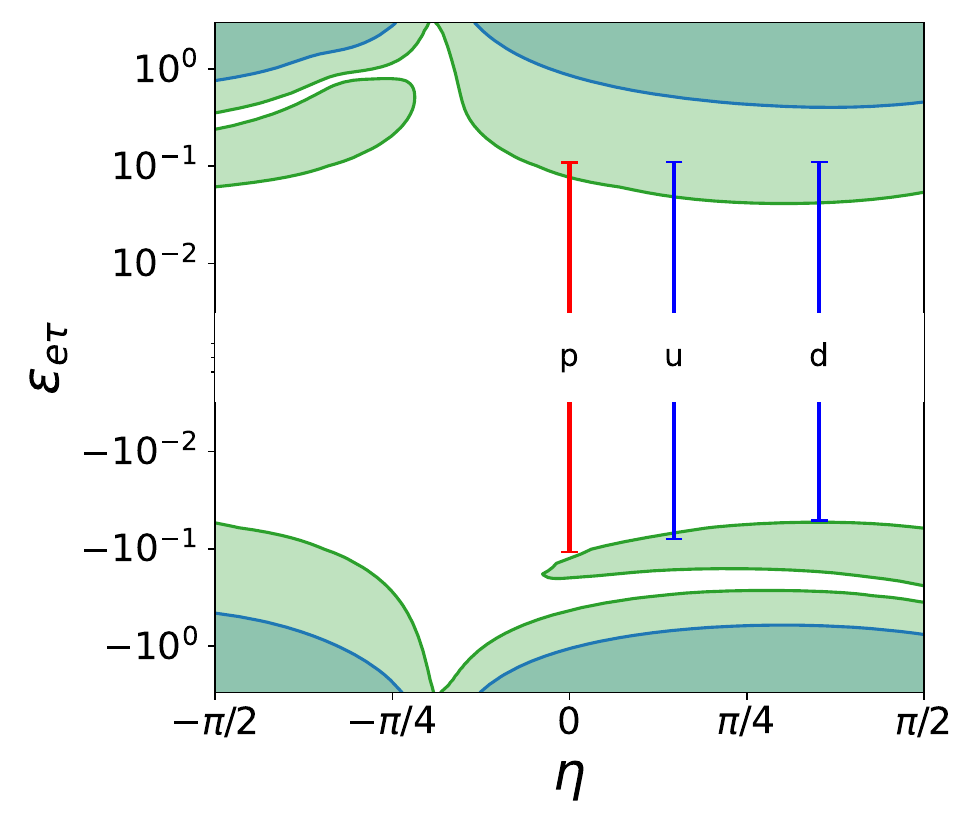} \\
\includegraphics[width=.49\textwidth]{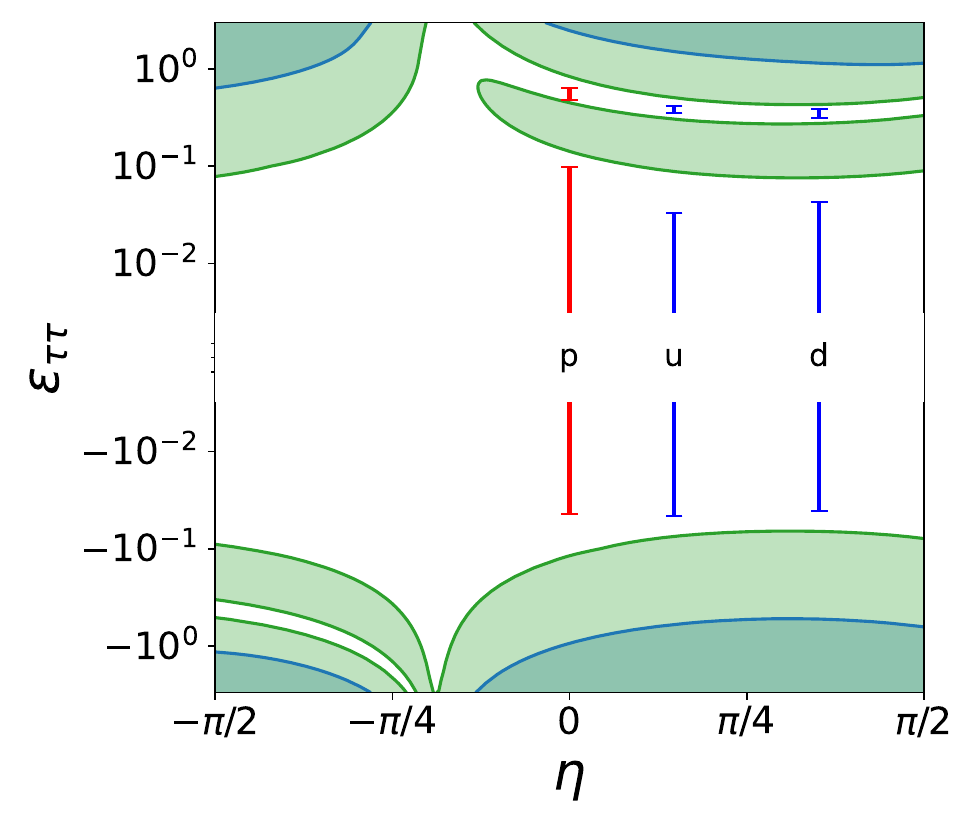} 
\includegraphics[width=.49\textwidth]{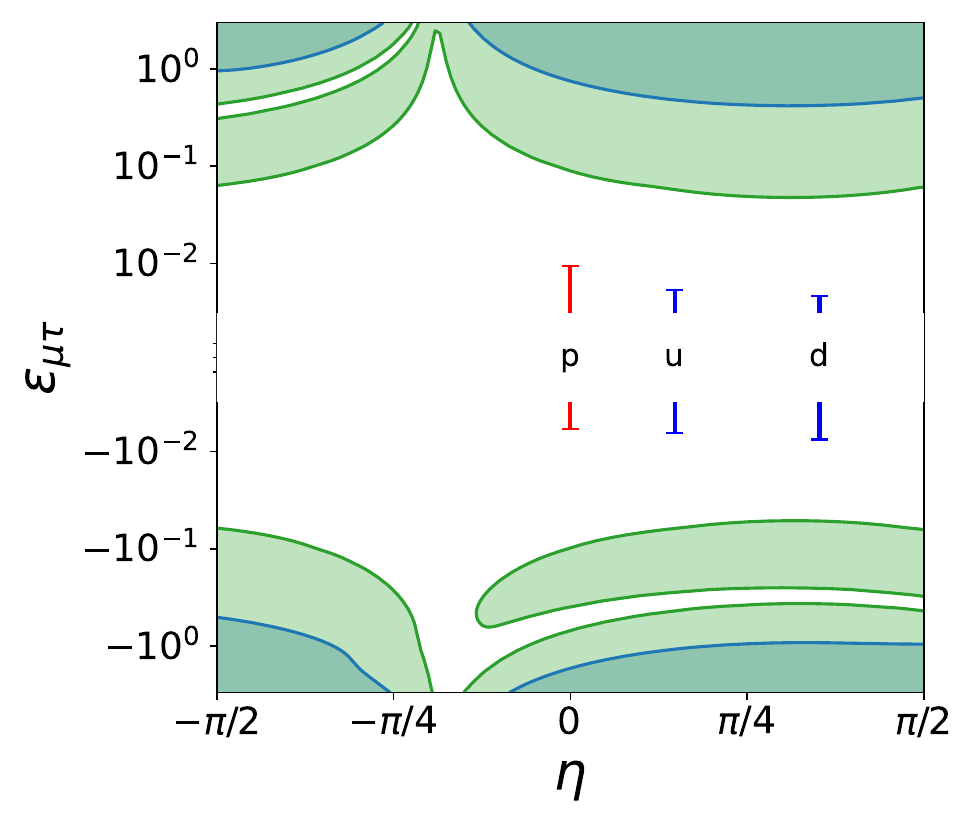}
\caption{The 90\% C.L.~sensitivity of the \RESNOVA{} experiment considering an exposure of 1~\tonyr and an energy threshold of 0.5~keV. 
The NSI parameters and the color coding are the same as in~\cref{fig:nsisensitivity_1kev_1ton}.}
\label{fig:nsisensitivity_0.5kev_1ton}
\end{figure}

\begin{figure}[htbp]
\centering
\includegraphics[width=.49\textwidth]{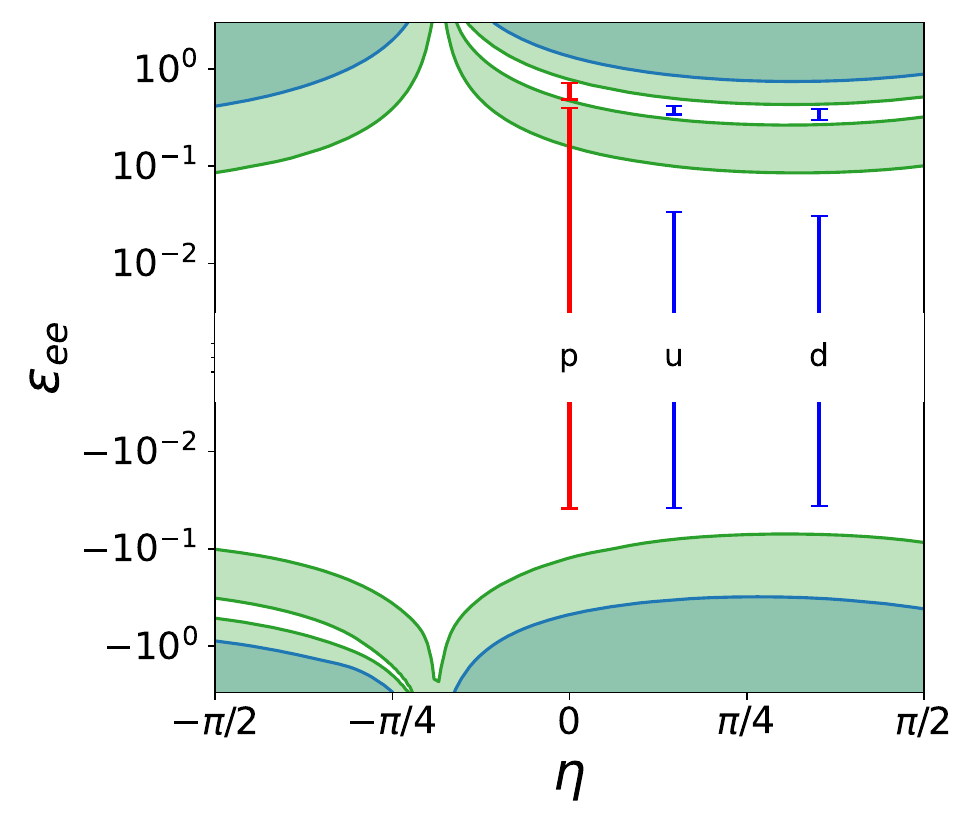}
\includegraphics[width=.49\textwidth]{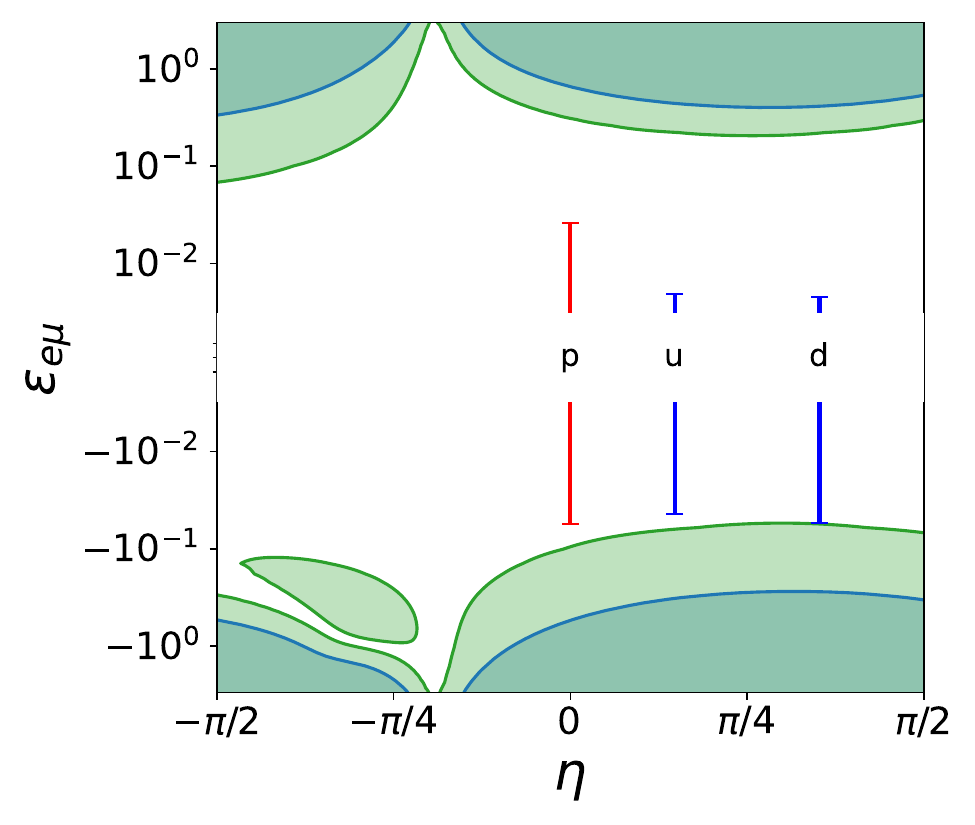} \\
\includegraphics[width=.49\textwidth]{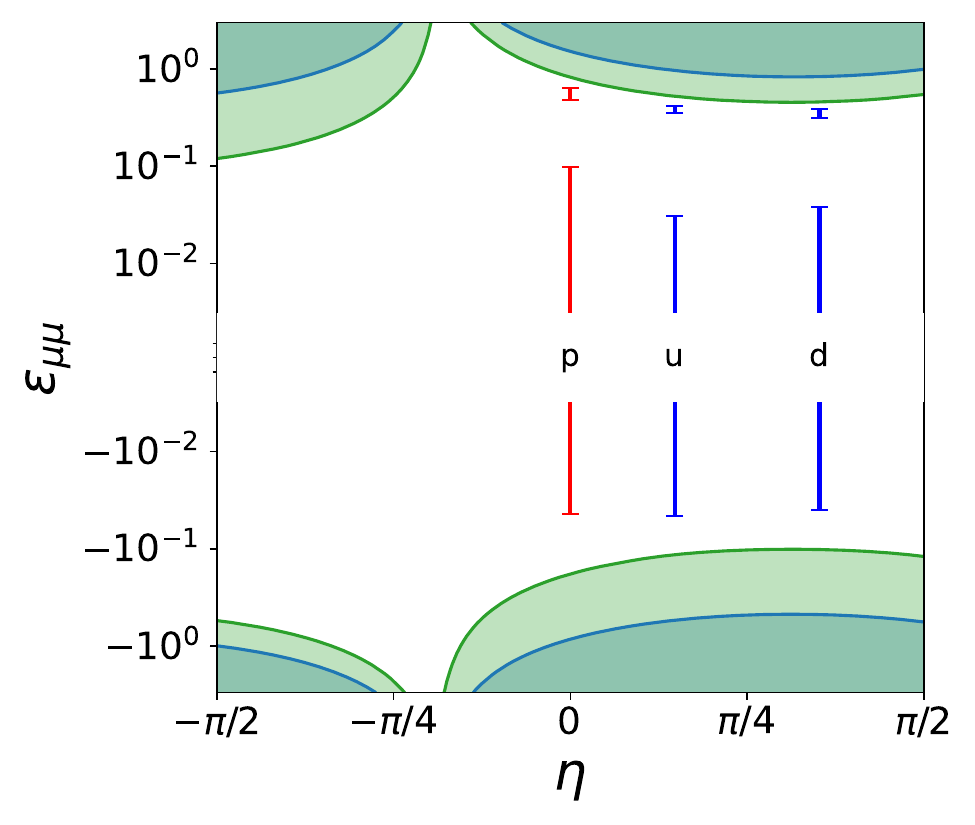}
\includegraphics[width=.49\textwidth]{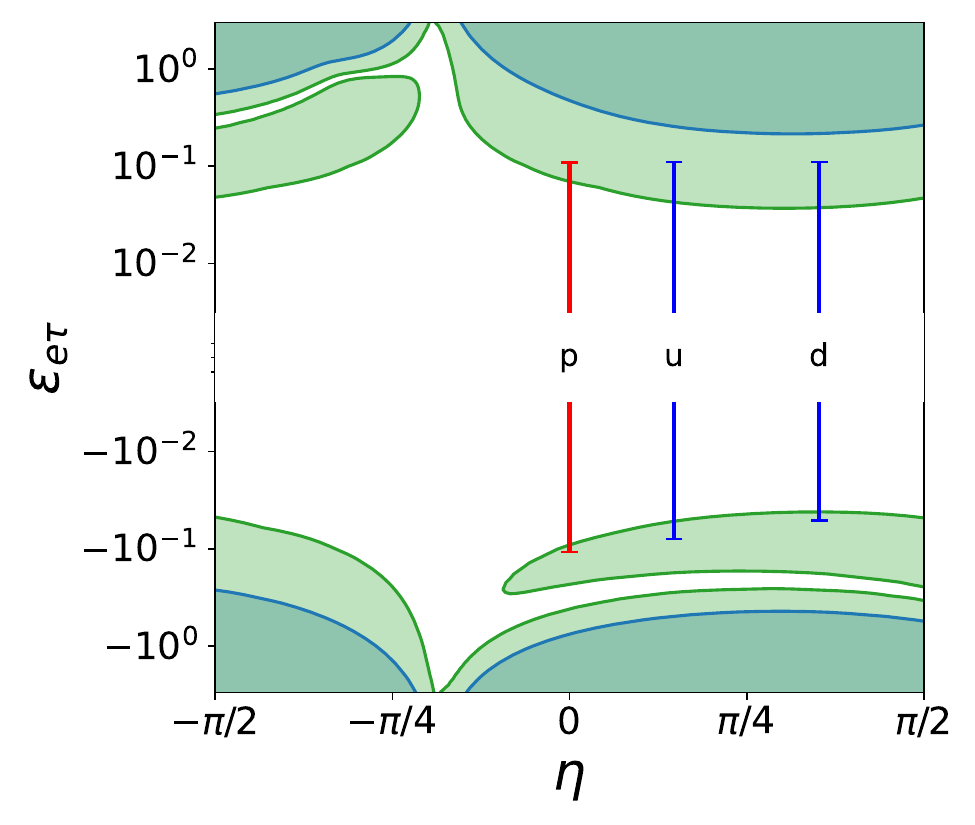} \\
\includegraphics[width=.49\textwidth]{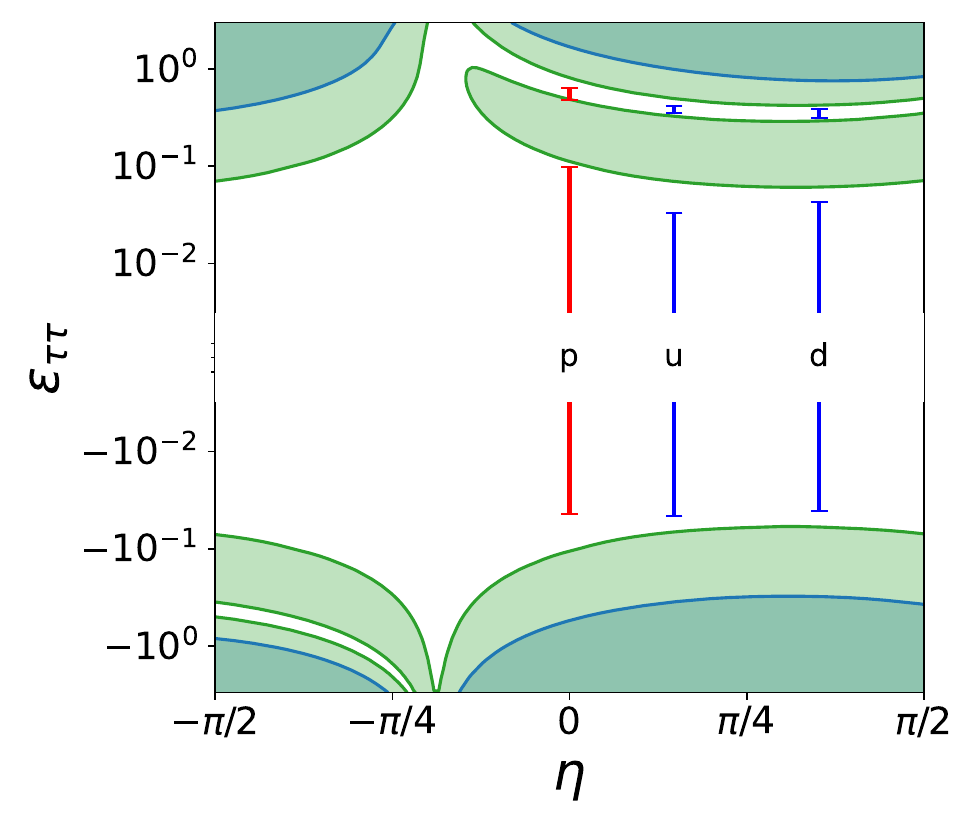}
\includegraphics[width=.49\textwidth]{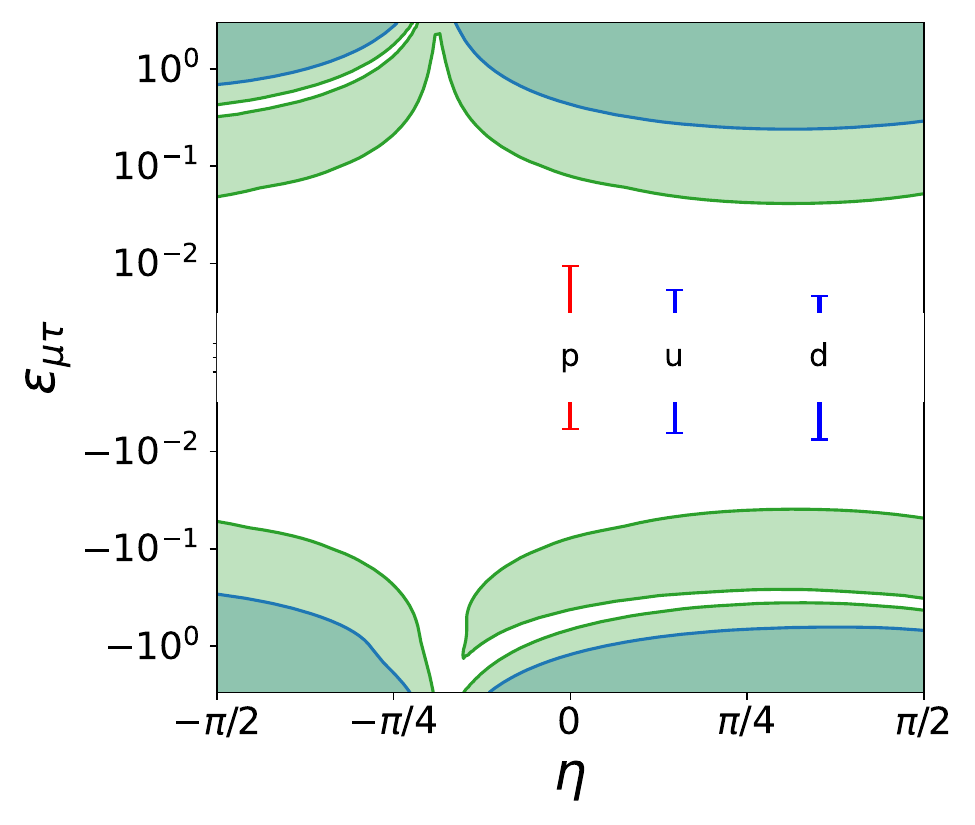}
\caption{The 90\% C.L.~sensitivity of the \RESNOVA{} experiment considering an exposure of 1~\tonyr and an energy threshold of 0.2~keV. 
The NSI parameters and the color coding are the same as in~\cref{fig:nsisensitivity_1kev_1ton}.}
\label{fig:nsisensitivity_0.1kev_1ton}
\end{figure}

\begin{figure}[htbp]
\centering
\includegraphics[width=.49\textwidth]{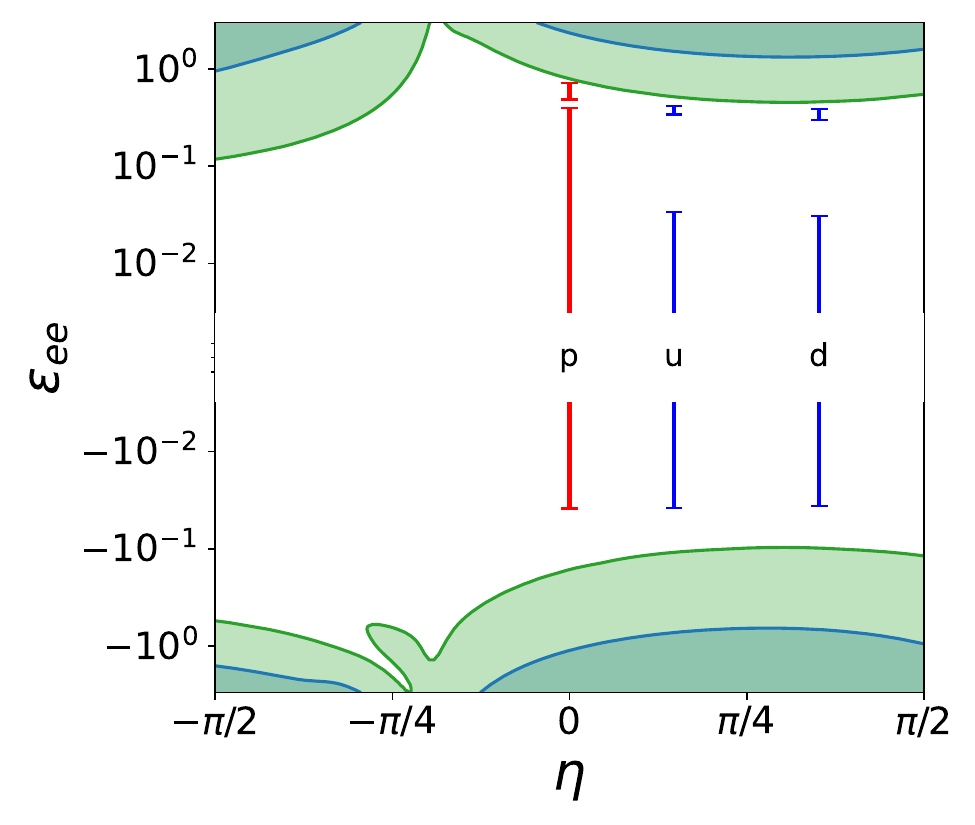}
\includegraphics[width=.49\textwidth]{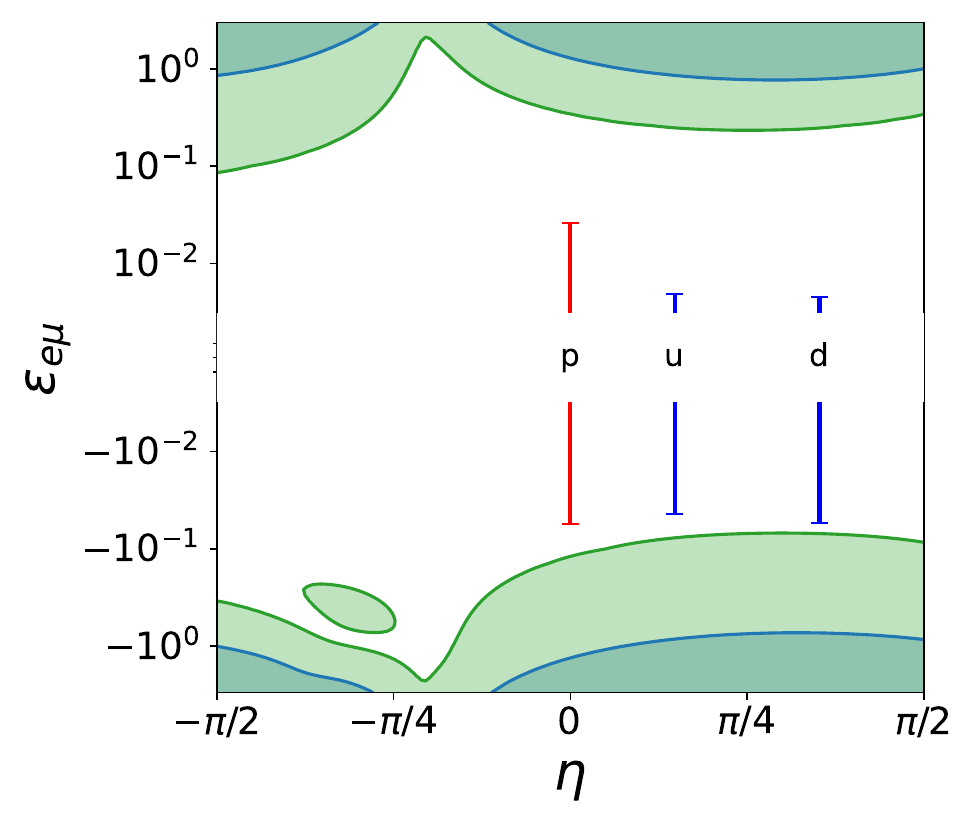} \\
\includegraphics[width=.49\textwidth]{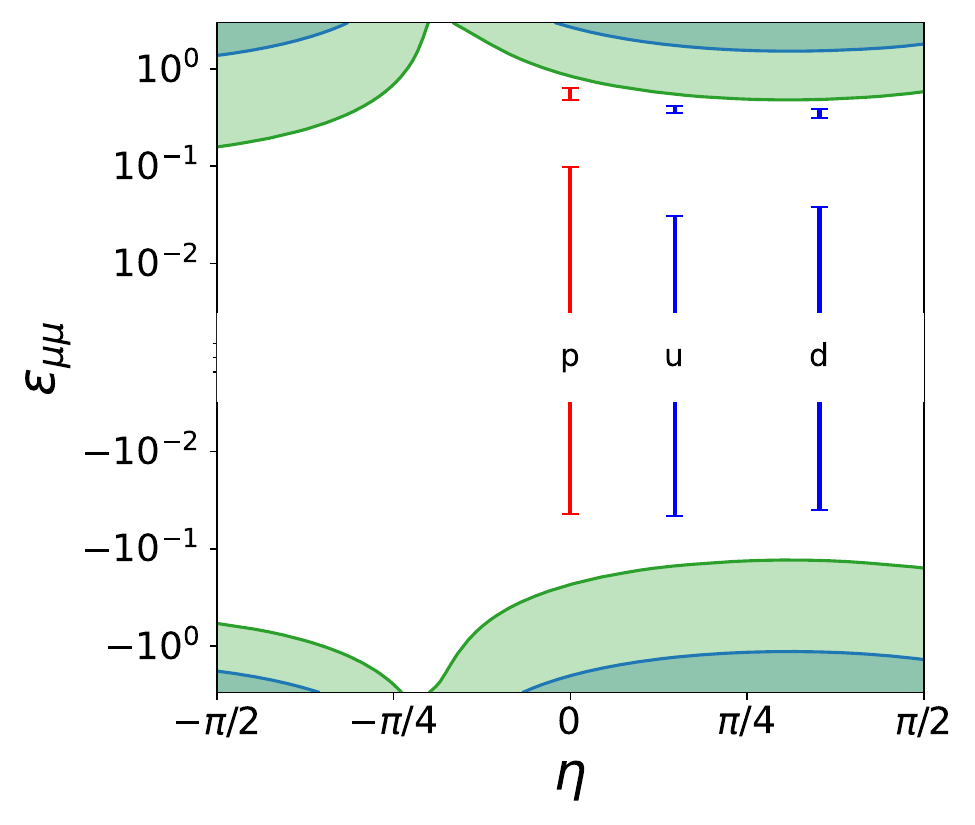}
\includegraphics[width=.49\textwidth]{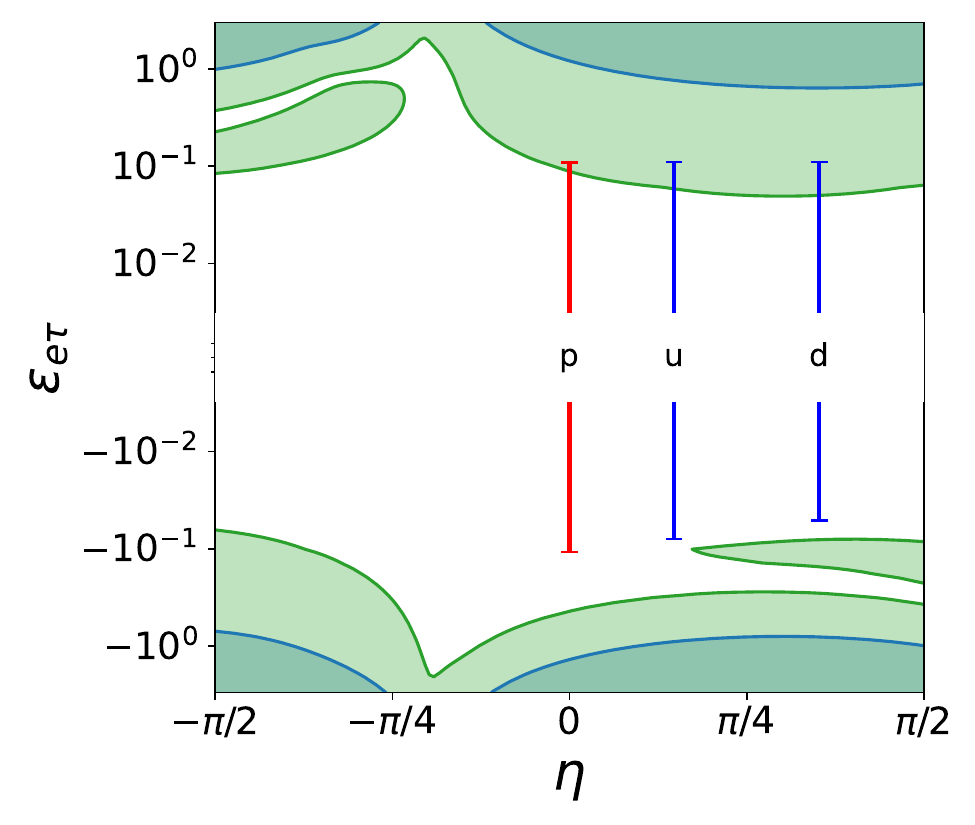} \\
\includegraphics[width=.49\textwidth]{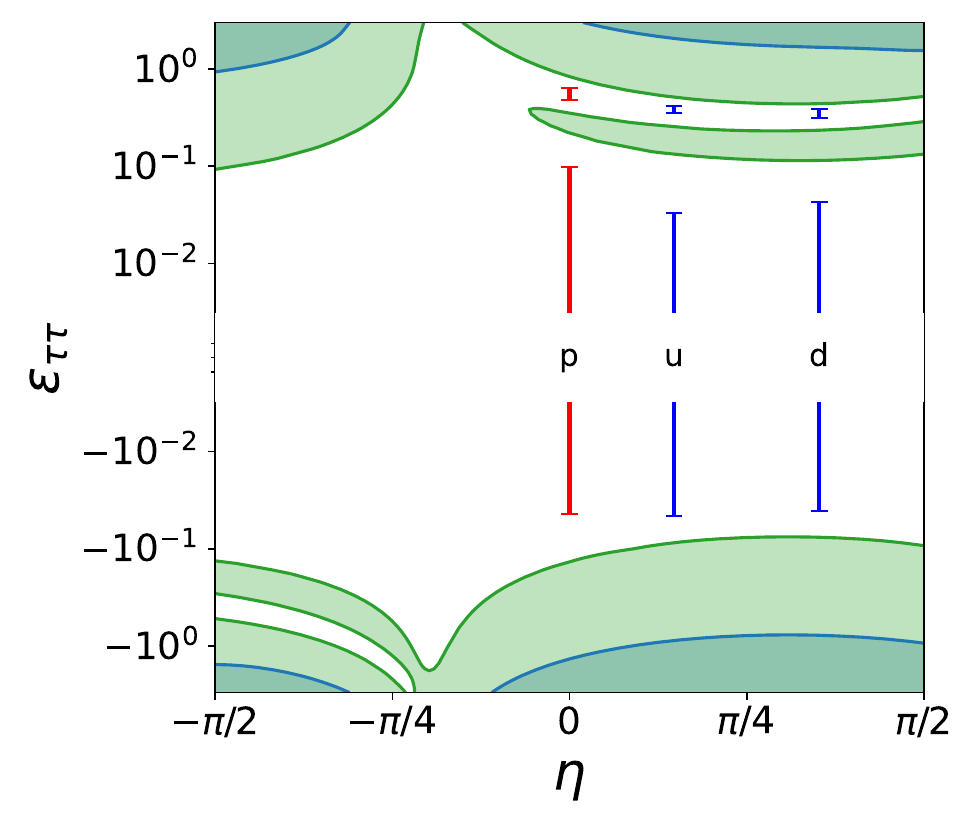}
\includegraphics[width=.49\textwidth]{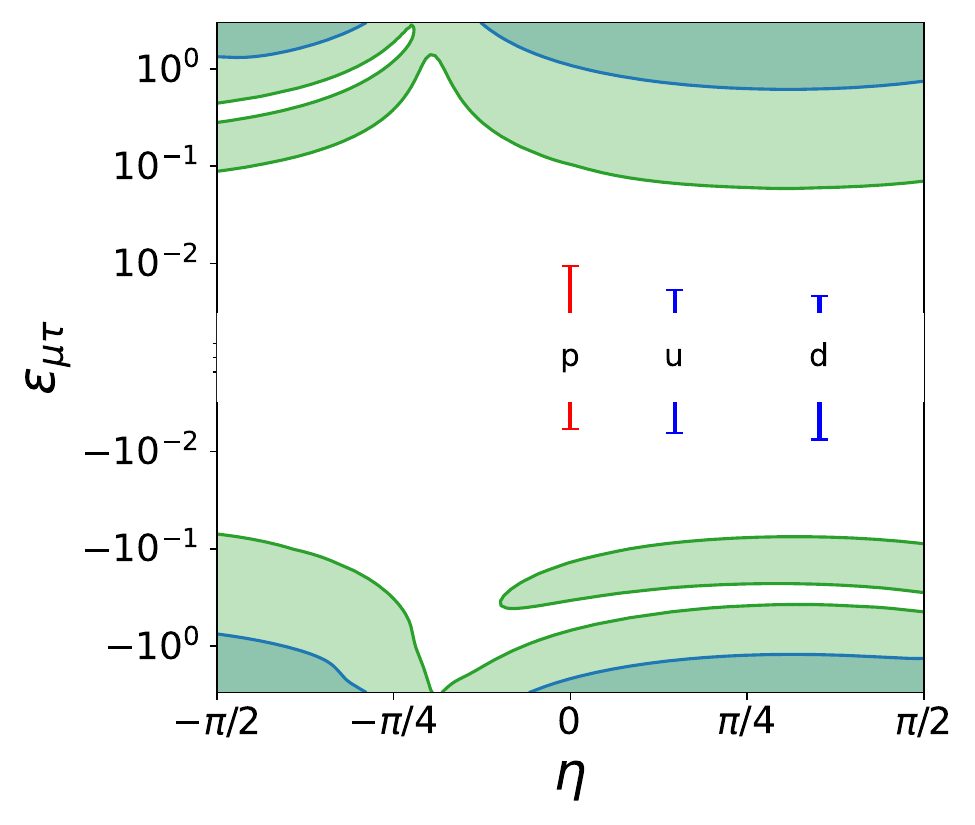}
\caption{The 90\% C.L.~sensitivity of the \RESNOVA{} experiment considering an exposure of 10~\tonyr and an energy threshold of 1~keV. 
The NSI parameters and the color coding are the same as in~\cref{fig:nsisensitivity_1kev_1ton}.}
\label{fig:nsisensitivity_1kev_10ton}
\end{figure}

In \cref{fig:nsisensitivity_0.1kev_1ton,fig:nsisensitivity_0.5kev_1ton,fig:nsisensitivity_1kev_10ton,fig:nsisensitivity_1kev_1ton}, we represent the projected sensitivity of different configurations of \RESNOVA{} on the neutrino NSI parameter space, as defined in \cref{eq:full_nsi}.
The green contours represent the sensitivities achieved in the optimal background rejection scenario, while the blue contours correspond to the worst-case scenario with no background rejection capability.%
\footnote{Note that the adiabatic approximation used to compute the neutrino evolution in the solar medium can break down for large values of the NSI couplings, $\varepsilon_{\alpha\beta}$. The regions of parameter space where adiabticity might break down has been investigated in Ref.~\cite{Amaral:2023tbs}
and is illustrated by the gray areas in Fig.~4 in that same reference. Since this happens mostly for large magnitudes of $|\varepsilon_{\alpha\beta}|\sim \mathcal{O}(1)$, where NSI are already excluded, we refrain from showing these areas in our~\cref{fig:nsisensitivity_1kev_1ton,fig:nsisensitivity_0.5kev_1ton,fig:nsisensitivity_0.1kev_1ton,fig:nsisensitivity_1kev_10ton}.}
The red bars denote the $2\,\sigma$ credible intervals from the global fits on NSI with only protons, $\varepsilon^p_{\alpha\beta}$, taken from Ref.~\cite{Coloma:2019mbs}, while the blue bars denote the same intervals with only up quarks, $\varepsilon^u_{\alpha\beta}$, and down quarks, $\varepsilon^d_{\alpha\beta}$, taken from Ref.~\cite{Coloma:2023ixt}.\footnote{Note that in order to translate the fit results on  to the radial variable $\varepsilon^{\eta,\varphi}_{\alpha\beta}$, the reported intervals have to be rescaled according to~\cref{eq:nsi_param} by the corresponding $\xi^f$. This results in a rescaling of the proton results by $\xi^p=\sqrt{5}$ according to $\varepsilon^{\eta,\varphi}_{\alpha\beta} = \varepsilon^p_{\alpha\beta}/\xi^p$, while by construction $\xi^u=1$ and $\xi^d=1$.}

The sensitivity reported in~\cref{fig:nsisensitivity_1kev_1ton} is already within reach of the \RESNOVA{} demonstrator currently under commissioning, provided a 6-year measurement campaign. With the full size experiment, the same would be in reach with a running time of 7 months, and we show in~\cref{fig:nsisensitivity_1kev_10ton} the projected sensitivity with a ten time increase in exposure.
\cref{fig:nsisensitivity_0.5kev_1ton} and~\cref{fig:nsisensitivity_0.1kev_1ton} investigate a minor and a more substantial improvement in the nuclear recoil detection energy threshold (0.5~keV and 0.2~keV, respectively). 
These projections highlight the importance of background rejection when it comes to reaching unchartered regions of the NSI parameter space. Furthermore, sensitivities computed assuming an improved energy detection threshold demonstrate that, at the same background level, a significantly larger portion of the parameter space would become accessible (\cref{fig:nsisensitivity_0.5kev_1ton,fig:nsisensitivity_0.1kev_1ton}), particularly for the $\nu_e$ and $\nu_\tau$ sector. Despite the technological feasibility, we do not investigate energy thresholds below 200~eV in order to avoid potential biases associated with the low-energy excess reported by several experiments in this regime~\cite{LEE}. By adopting a conservative threshold of 200~eV, we ensure that our sensitivity projections remain robust against unresolved detector effects, background mismodelling, and the current experimental uncertainties that dominate at sub-200~eV energies.

From~\cref{fig:nsisensitivity_1kev_1ton}, we can see that the demonstrator version of \RESNOVA{} with a nominal exposure of 1 \tonyr and a threshold of 1 keV  will not yet be sensitive enough to further constrain the currently allowed NSI intervals from global fits. Yet, we can see that in case of optimal background rejection (green contour), \RESNOVA{} will be able to place similarly stringent bounds on flavor-changing $e \tau$ NSI.

In~\cref{fig:nsisensitivity_0.5kev_1ton,fig:nsisensitivity_0.1kev_1ton}, however, we can see that the sensitivity of \RESNOVA{} to NSI in \cevns increases already significantly for a moderate improvement in the energy threshold down to 0.5 keV, and even more for a substantial improvement in threshold down to 0.2~keV. The diminishing difference between the sensitivities obtained under pessimistic and optimistic background assumptions at lower energy thresholds can be understood in terms of the relative scaling of signal and background rates. Lowering the threshold increases the expected signal more rapidly than the background, so that the discovery reach improves in both cases. Consequently, at sufficiently low thresholds the 90\% C.L.~sensitivities for the high- and low-background assumptions tend to converge. This explains why the impact of background modeling on the projected sensitivity becomes progressively less pronounced as the threshold decreases. In this sense, threshold improvements act more effectively than an increased exposure. Therefore, comparable sensitivity may be achieved with reduced exposure if sufficiently low thresholds are realized, while concurrent improvements in both parameters would provide a synergistic gain in reach.

With lower thresholds, \RESNOVA{} has the potential to improve the constraints over the global fits on $e\tau$ flavor-changing NSI in the proton, up-quark and down-quark projection. Moreover, \RESNOVA{} can improve over the global fits of flavor-diagonal NSI with protons, both in the $ee$- and $\tau\tau$-direction, albeit rather modestly. Conversely, if we consider a larger detector with an exposure of 10~\tonyr, but the nominal threshold of 1~keV, we can see in~\cref{fig:nsisensitivity_1kev_10ton} that we can expect similar improvements in sensitivity as for lowering the threshold down to 0.5~keV.

The sensitivities of~\cref{fig:nsisensitivity_0.5kev_1ton,fig:nsisensitivity_0.1kev_1ton,fig:nsisensitivity_1kev_10ton} are comparable to those  that xenon experiments can achieve with an exposure of $\mathcal{O}(20)$ \tonyr~\cite{Amaral:2023tbs}. 
Due to the enhancement of the coherence factor in the \cevns cross section with the larger mass number of lead compared to xenon, \RESNOVA{} will detect a similar number of events at more modest exposures of $\mathcal{O}(1)$ \tonyr.
Similarly, the material-dependent blind direction occurring for $\eta=\tan ^{-1}(-{Z}/{N}\, \cos \varphi)$~\cite{Amaral:2023tbs},
for which the effects of proton-/electron- and neutron-NSIs cancel, appear at a similar angle of $\eta\approx -{3}/{16}\,\pi$ compared to xenon-based experiments. This is expected, since Pb and Xe have a similar proton-to-neutron ratio, $Z/N$.

It is interesting to note that \RESNOVA{} could potentially probe the region of parameter space in $\varepsilon_{e\mu}$ and $\varepsilon_{e\tau}$ that could resolve the tension between the data of the T2K and NOvA experiments for NSI magnitudes of $|\varepsilon_{\alpha\beta}|\approx0.2$ and a NSI phase of $\phi_{\alpha\beta}\approx 3\pi/2$, as  
pointed out in Ref.~\cite{Denton:2020uda}.\footnote{We would like to thank Peter Denton for pointing this out to us.} 
The required NSI magnitudes could be tested under several experimental configurations: (i) with a 10~\tonyr exposure in the optimistic background scenario; (ii) with a 1~\tonyr exposure assuming a moderate improvement in the energy threshold in the optimistic background scenario; and (iii) with a 1~\tonyr exposure assuming a substantial improvement in the energy threshold even in the pessimistic background scenario.

These results show that cryogenic bolometers of tonne-scale target masses, like \RESNOVA{}, will become serious contenders in the quest to constrain BSM neutrino physics. This strengthens the physics case for these types of detectors as not only pure DM experiments, but also as novel observatories for astrophysical neutrinos.

\section{Conclusions}
\label{sec:summary}

In this work, we have investigated the potential of cryogenic PbWO$_4$ detectors to probe neutrino non-standard interactions through coherent elastic neutrino--nucleus scattering of solar neutrinos. 
Building upon the NSI formalism developed in Refs.~\cite{Coloma:2019mbs,Coloma:2022umy,Amaral:2023tbs,Coloma:2023ixt}, we have focused on the nuclear recoil channel in a heavy, high-$A$ target, assessing the sensitivity of the \RESNOVA{} detector concept under realistic assumptions on energy thresholds, exposures, and background levels.

Thanks to the large neutron number of Pb and W nuclei, CE$\nu$NS in PbWO$_4$ benefits from a substantial coherence enhancement, making \RESNOVA{} sensitive to NSI scenarios that modify neutrino--nucleus interactions. As a neutral-current process, CE$\nu$NS provides a flavor-independent probe of solar neutrinos, complementing traditional solar neutrino experiments based on electron scattering and oscillation measurements. In particular, direct detection experiments are uniquely sensitive to \cevns with $\nu_\tau$. 

Our results in~\cref{fig:thr_vs_exp} show that, for the benchmark \RESNOVA{} configuration with an energy threshold of $\mathcal{O}(1~\mathrm{keV})$, the expected signal from standard solar neutrino fluxes lies close to the experimental sensitivity frontier, detectable at the $3\sigma$ level with exposures of $\epsilon\sim\mathcal{O}(50)$ \tonyr in the optimistic background scenario. In this regime, the experiment is not expected to observe the Standard Model solar neutrino signal in a conventional detection scenario, but remains sensitive to deviations induced by NSI, which manifest as an excess in the nuclear recoil spectrum. For exposures at the \tonyr scale, sizable regions of the NSI parameter space can therefore be probed even when the SM contribution is marginally accessible.

The achievable sensitivity is strongly driven by the detector energy threshold and by the control of intrinsic radioactive backgrounds in the crystal bulk. While the baseline \RESNOVA{} design targets a threshold of approximately 1~keV, we have also explored more favorable scenarios with thresholds of 0.5~keV and 0.2~keV, which could be achieved by exploiting detector granularity and trading target mass for an increased number of readout channels. In such conditions, the detection of solar neutrinos via SM CE$\nu$NS will become accessible.

Overall, our study demonstrates that cryogenic PbWO$_4$ detectors constitute a promising and complementary approach to the exploration of the neutrino NSI landscape. \RESNOVA{} has the potential to bridge the gap between conventional neutrino experiments and direct detection searches, providing sensitivity to new physics effects in the low-energy neutrino sector. In the context of NSI, a next step would be to investigate whether \cevns measurements at \RESNOVA{} could help to further break the degeneracies in determining the neutrino mass ordering associated with the LMA-Dark solution.
The forthcoming \RESNOVA{} demonstrator will be a crucial step toward validating the required background and threshold performance, and toward assessing the feasibility of a scalable, flavor-independent solar neutrino observatory based on CE$\nu$NS.

\acknowledgments

This work received funding from the EU Horizon Europe program through grant ERC-101087295-RES-NOVA.
We are grateful to the University of Milano-Bicocca (UNIMIB) and the Istituto Nazionale di Fisica Nucleare (INFN) for actively supporting the collaboration.
DC and PF acknowledge support from the Spanish Agencia Estatal de Investigaci\'on through the grants PID2024-155874NB-C22 and CEX2020-001007-S, funded by MCIN/AEI/10.13039/501100011033. AC acknowledges the support of S. Ge, funded by the National Natural Science Foundation of China (Grant Nos. 12425506, 12375101, 12090060, and 12090064) and the Double First-Class startup funds by Shanghai Jiao Tong University. The work of PF was supported by a fellowship from La Caixa Foundation (ID 100010434 with code LCF/BQ/PR25/12110019). SG acknowledges support from the U.S. Department of Energy (DOE) Grant No. [DE-SC0011091].
FAD and VIT acknowledge the support of Laboratori Nazionali del Gran Sasso and Gran Sasso Science Institute during the difficult period in Ukraine.


\bibliographystyle{JHEP}
\bibliography{biblio.bib}

\end{document}